%% file: bare_jrnl_new_sample4.tex
\documentclass[lettersize,journal]{IEEEtran}
\usepackage{amsmath,amsfonts}
\usepackage{algorithmic}
\usepackage{algorithm}
\usepackage{array}
\usepackage[caption=false,font=normalsize,labelfont=sf,textfont=sf]{subfig}
\usepackage{textcomp}
\usepackage{stfloats}
\usepackage{url}
\usepackage{verbatim}
\usepackage{graphicx}
\usepackage{cite}
\usepackage{multirow}
\usepackage[table]{xcolor}
\usepackage{physics}
\begin{document}

\title{Pedestrian and Passenger Interaction with Autonomous Vehicles: Field Study in a Crosswalk Scenario}

\author{Rubén Izquierdo, Javier Alonso, Ola Benderius, Miguel Ángel Sotelo, David Fernández Llorca
        % <-this % stops a space
\IEEEcompsocitemizethanks{
\IEEEcompsocthanksitem R. Izquierdo, J. Alonso, M. Á. Sotelo and D. Fernández Llorca, are with the Computer Engineering Department, University of Alcalá, Alcalá de Henares, Madrid, Spain. \protect \IEEEcompsocthanksitem O. Benderius is with the Department of Mechanics and Maritime Sciences, Chalmers University of Technology, Gothenburg, Sweden.\protect \IEEEcompsocthanksitem D. Fernández Llorca is with the Joint Research Centre, European Commission, Sevilla, Spain. \protect\\ 
E-mail: ruben.izquierdo@uah.es; david.fernandez-llorca@ec.europa.eu.
}
%\thanks{Manuscript received November, 2023; revised MMMM DD, 2024.}
}

\markboth{Journal of \LaTeX\ Class Files,~Vol.~xx, No.~yy, December~2023}%
{Shell \MakeLowercase{\textit{et al.}}: A Sample Article Using IEEEtran.cls for IEEE Journals}

%\IEEEpubid{0000--0000/00\$00.00~\copyright~2023 IEEE}

\maketitle

\begin{abstract}
%This paper presents the results of real-world testing of human-vehicle interactions with an autonomous vehicle equipped with internal and external Human Machine Interfaces (HMIs) in a crosswalk scenario. The internal and external HMIs were combined with implicit communication techniques using gentle and aggressive braking maneuvers in the crosswalk. 
%Results have been collected in the form of questionnaires and measurable variables such as distance or speed when the pedestrian decides to cross.
%The questionnaires show that pedestrians feel safer when the external HMI or the gentle braking maneuver is used interchangeably, while the measured variables show that external HMI only helps in combination with the gentle braking maneuver.
%The questionnaires also show that internal HMI only improves passenger confidence in combination with the aggressive braking maneuver.

This study presents the outcomes of empirical investigations pertaining to human-vehicle interactions involving an autonomous vehicle equipped with both internal and external Human Machine Interfaces (HMIs) within a crosswalk scenario. The internal and external HMIs were integrated with implicit communication techniques, incorporating a combination of gentle and aggressive braking maneuvers within the crosswalk. Data were collected through a combination of questionnaires and quantifiable metrics, including pedestrian decision to cross related to the vehicle distance and speed. The questionnaire responses reveal that pedestrians experience enhanced safety perceptions when the external HMI and gentle braking maneuvers are used in tandem. In contrast, the measured variables demonstrate that the external HMI proves effective when complemented by the gentle braking maneuver. Furthermore, the questionnaire results highlight that the internal HMI enhances passenger confidence only when paired with the aggressive braking maneuver.

\end{abstract}


\input{1_introduction}
\input{2_arte}
\input{3_description}
\input{4_results}

\input{5_discussion}
\input{6_conclusions}

\section*{Disclaimer}
The views expressed in this article are purely those of the authors and may not, under any circumstances, be regarded as 
an official position of the European Commission. 

\section*{Acknowledgments}
This work was mainly funded by the HUMAINT project by the Directorate-General Joint Research Centre of the European Commission. It was also partially funded by Research Grants PID2020-114924RB-I00 and PDC2021-121324-I00 (Spanish Ministry of Science and Innovation).

% {\appendix[Proof of the Zonklar Equations]
% Use $\backslash${\tt{appendix}} if you have a single appendix:
% Do not use $\backslash${\tt{section}} anymore after $\backslash${\tt{appendix}}, only $\backslash${\tt{section*}}.
% If you have multiple appendixes use $\backslash${\tt{appendices}} then use $\backslash${\tt{section}} to start each appendix.
% You must declare a $\backslash${\tt{section}} before using any $\backslash${\tt{subsection}} or using $\backslash${\tt{label}} ($\backslash${\tt{appendices}} by itself
%  starts a section numbered zero.)}

\appendices
\section{Study Questionnaire 1}\label{appendix:questions1}
\input{8_anexo_I}
\section{Study Questionnaire 2}\label{appendix:questions2}
\input{8_anexo_II}

% \section{References Section}
% You can use a bibliography generated by BibTeX as a .bbl file.
%  BibTeX documentation can be easily obtained at:
%  http://mirror.ctan.org/biblio/bibtex/contrib/doc/
%  The IEEEtran BibTeX style support page is:
%  http://www.michaelshell.org/tex/ieeetran/bibtex/
 

\bibliographystyle{IEEEtran}
\bibliography{citations.bib}

\include{7_bio.tex}

\end{document}

%% file: 1_introduction.tex
\section{Introduction}\label{sec:intro}
%\IEEEPARstart{T}{his} file is intended to serve as a ``sample article file''.

\IEEEPARstart{T}{rustworthy} human-vehicle interaction in the context of Autonomous Vehicles (AVs) 
%\footnote{We use Autonomous Vehicle to refer to Highly Automated or Driverless vehicles, where users inside the vehicle are mere passengers.} 
has a fundamental impact on the user's sense of agency, perception of risk, and trust \cite{Li2019}. These factors, in turn, are essential to avoid both disuse and misuse of technology, which directly affect user acceptance and safety respectively \cite{Llorca2023}.

Human-vehicle interaction in autonomous driving is a multi-user problem that primarily involves two groups of people: those using the AV (passengers) and external road users interacting with the AV (i.e., pedestrians, cyclists, drivers). The absence of a driver to communicate with, from both the perspective of a passenger and an external road agent, alters the nature and dynamics of interactions \cite{Detjen2021, Rasouli2020}. In this new context, AVs need to communicate their intentions to road agents that are not automated or connected, such as regular vehicles, pedestrians, or cyclists, in the same way that regular drivers convey their intentions using visual cues or the vehicle dynamics itself. This communication process becomes especially crucial in scenarios where safety-relevant interactions may occur, such as when a pedestrian is crossing the road in front of a vehicle.

\begin{figure}
%\centerline{\includegraphics[width=\columnwidth]{figures/Abstract.pdf}}
\centerline{\includegraphics[width=\columnwidth]{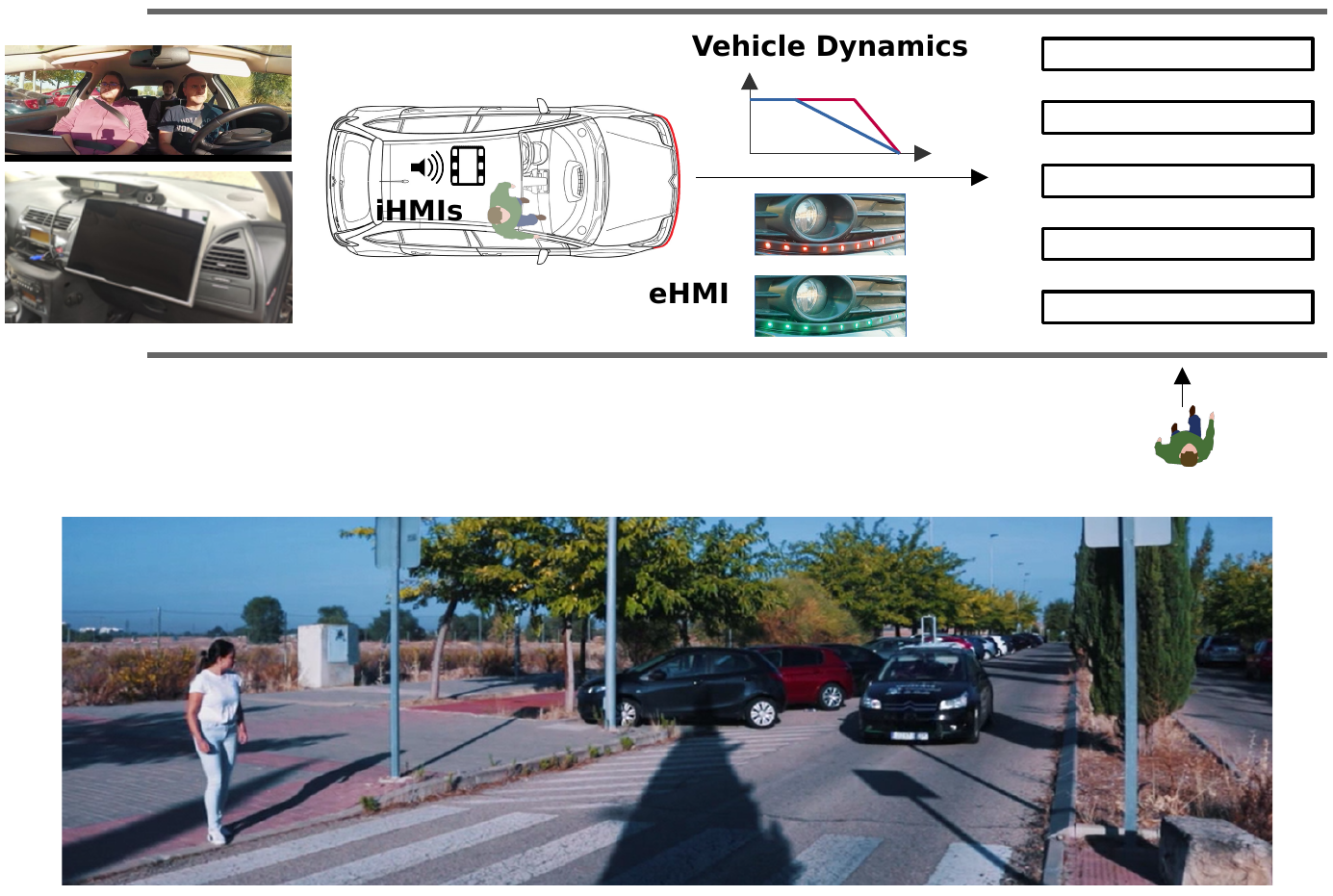}}
\caption{Top: schematic overview of the experiment. Bottom: actual image of the field test scenario.}
\label{fig:schematic1}
\end{figure}

The use of Vehicle-to-Everything (V2X) technology \cite{ignacio_v2v} facilitates communication between other automated agents, such as other connected vehicles and infrastructure, but it still leaves humans unaware of the vehicle's intentions. Human-vehicle interaction primarily occurs through human-machine interfaces (HMIs), both internal (iHMI) and external (eHMI). The specific modality of these interfaces is tied to vehicle technology and human capabilities \cite{Llorca2023}. The behavior of the vehicle, i.e., its movement dynamics, also serves as an important form of implicit communication with a significant impact on the interaction \cite{Rasouli2020}, \cite{dey2021communicating}.

The impact of these forms of explicit or implicit communication on in-vehicle users (drivers or passengers) and other external road users has been widely studied, but always separately, which prevents drawing holistic conclusions. From an experimental perspective, previous work has focused on simulated environments using virtual reality \cite{Martin2023}, or on real environments with two main types of constraints. On one hand, there are cases in which the pedestrian only expresses an intention to cross without actually performing the crossing action \cite{dey2021communicating}. On the other hand, there are cases where the driving is not truly automatic but mediated by Wizard of Oz methods \cite{Lagstrom2015}. In all cases, the results are somewhat limited due to the mismatch with real-world interaction scenarios.

In this work, we present the results of a real field study on human-vehicle interaction in crosswalk scenarios, involving both pedestrians and passengers. We expand upon our preliminary study \cite{Izquierdo2023} by providing more details about the experimental setup, data gathered, results, and discussion. Our automated test vehicle \cite{drivertive2018} is not mediated by the Wizard of Oz approach. The pedestrians do not explicitly communicate their intention to cross (which adds the difficulty of identifying the exact moment when the pedestrian decides to cross), but decide to cross or not, and complete their crossing action naturally. This approach allows us to draw conclusions that consider both types of users who interact with the AV in a holistic manner. It also enables us to investigate the impact of previous interaction experience as a passenger on pedestrian behavior and vice versa. Furthermore, we can minimize the gap between the interactions measured in our experimental setup and those that would occur in a real environment. We evaluate different types of internal and external HMIs, as well as implicit communication through vehicle dynamics, using both behavioral and attitudinal evaluation methods.

%% file: 2_arte.tex
\section{Related work}

The interaction between passengers and autonomous vehicles is a rapidly evolving and multifaceted field that encompasses various aspects of technology, psychology, and human-vehicle interfaces.
Initially, the interaction between vehicles and their drivers and passengers has been studied in depth for many years by car-makers. In this way, the traditional dashboard has become a sophisticated piece of software and hardware to transmit a large amount of information to the driver regarding the state of the vehicle and the environment surrounding the car in the last decade\cite{berkery2004future}.

The automation level is a key point in the study of the interaction. For low levels of SAE automation \cite{SAE} (L1 to L3) the interaction between vehicles and passengers has been limited to infotainment systems as drivers are in charge of the driving task and passengers are not. 
%In the new AV paradigm, for SAE automation levels 1-3 a driver or a backup driver is needed for the driving task, 
However, for higher automation levels (L4 and L5) there are no longer any drivers, they become users or passengers. The users of the AV service can neglect the driving task, which is associated with higher driving comfort \cite{engelbrecht2013fahrkomfort}. Another, more fundamental and unfamiliar, aspect is the allocation of an automated system in contrast to a human driver. To compensate for the lack of a human driver to rely on, the AV must provide some information related to the driving task to the passengers even if they are not actively supervising the driving task.
%However, the lack of a human driver to rely on has to be compensated with some kind of information from the VA.

Internal HMIs have the mission to interact with users to reduce the stress \cite{elbanhawi2015passenger}, anxiety \cite{baker1992discomfort}, and perception of risk that can arise when the vehicle does not behave as the passenger expects \cite{allsop2017eye}. In this way, the use of screens to share pictograms combined with audio messages is the mainstream option \cite{hartwich2020passenger}. The iHMI has the advantage of being integrated inside the vehicle cabin and can communicate information to passengers through visual and/or auditory clues \cite{benderius2017best}. A 16-inch screen with audio capability has been used in this work to interact with passengers.

The interaction of the vehicle with other external agents has been limited to the driving indicators such as blinkers and braking lights. Other signals like flashing lights or driving gestures are also used but they have non-standard meanings and can increase the potential risk of an interaction if they are misinterpreted. 
%Although some studies argue that one way to abuse drivers is to not establish eye contact and force the vehicle to stop and avoid an accident \cite{cita_de_Olaverri}, 
Although some studies argue that pedestrians make the crossing decision based solely on the vehicle's kinematics, without the need to establish eye contact with the driver \cite{AlAdawy2019}, 
others show that establishing some form of communication between the driver and the external agent is the best way to avoid an accident and increase the confidence in the interaction \cite{de2019perceived}.
Again, with the highest levels of automation (4 and 5) the external agents cannot establish eye contact with the driver because all the people inside the vehicle are mere passengers or the vehicle can be empty. The use of eHMIs is intended to replicate the traditional V2V communications (i.e. light codes) but extended to all types of road users, especially for front-of-vehicle agents. The efforts to share information from AVs have developed several topologies with different features \cite{bazilinskyy2019survey}: anthropomorphic, textual, light patterns or trajectory projection on the ground \cite{chang2022can, modes5, mason2022lighting}. The color and shape of the communication play a fundamental role in the interaction. Messages using red color are more suitable to indicate risk, and green ones to indicate a safe situation \cite{bazilinskyy2019survey}. This idea reinforces the theories of social constructivism \cite{lynch2016social}. However, in \cite{dey2021communicating} a turquoise LED strip is used to convey different AV behaviors varying the light pattern but not the color. Turquoise color is commonly used as a non-related-to-traffic color for "new" AV features. The most promising kind of eHMIs are light-based. Carmakers \cite{toyotaeHMI} and researchers \cite{rouchitsas2019external, habibovic2019external} are concentrating their efforts on this technology. Also, policymakers are more willing to approve this kind of new light features as they can follow similarities with current light-communication codes on regular vehicles \cite{heidi_project}. In this work, an RGB LED strip alongside the front bumper performing solid red and solid green patterns is used to interact with the pedestrian.

Studies used to assess human behavior in traffic scenes are mainly based on surveys \cite{wilde1980immediate, price2000relationship} or direct interviews \cite{crundall1999driving}. These forms of studies, however, have been criticized for the bias people have in answering questions, and also the quality of the questions asked in order to accurately collect the desired information\cite{dijksterhuis2015impact}. 
In addition, behavior can be analyzed via on-site observation by the researcher either present in the vehicle \cite{risser1985behavior} or standing outside \cite{tom2011gender}. 
Naturalistic recording of the traffic interaction (both videos and images \cite{kotseruba2016joint,dipietro1970pedestrian}), is one of the most effective methods for studying traffic behavior. In this method of study, a camera is placed either inside or outside the vehicle \cite{neale2005overview,rasouli2017agreeing} or outside on roadsides \cite{sun2003modeling, wang2010study}. Recording-based studies allow data to be obtained by direct measurement and allow certain biases to be eliminated \cite{tom2011gender}. New technological developments in combination with the use of AVs login systems endow different modalities of recording traffic events, such as eye-tracking \cite{clay1995driver, dey2021communicating} and positioning systems to measure the interaction in a distance and speed-based domain. Eye-tracking \cite{dey2021communicating} can be used to analyze the focus of attention of the study subject and which factors are most relevant. In this study, both questionnaires and internal and external video recordings are used to analyze interactions. In this work, surveys, recordings, and direct measures are used to study the interaction between the AV and the passenger and pedestrian.

Another important element in AV interaction studies is the vehicle itself and the existence of a person in the driver's seat. In the context of autonomous driving research, the Wizard of Oz technique \cite{lagstrom2016avip} is commonly used to replicate the behavior of an intelligent system but its use is limited to interaction with agents outside the vehicle. The ideal scenario is an empty driver's seat, but sometimes, for technical or legal requirements this is not possible. In this study, there is a backup driver in the driver's seat but there were no interactions between the backup driver and the vehicle during any of the experiments. Moreover, the backup driver is hidden behind the sun visor.

Current studies of AV are very limited and most of them do not perform real interactions\cite{martin2023digital, zou2023pedestrian}. In \cite{dey2021communicating} a Wizard of Oz study analyzed the interaction between an AV and a pedestrian in a crosswalk but the pedestrian never crossed in front of the vehicle, totally removing the actual exposure to risk with the already exposed problem \cite{Li2019}. The study concluded that the use of an eHMI and a slow pace of driving contribute to an increase in the pedestrian's willingness to cross.

%The iHMI has the advantage of being integrated inside the vehicle cabin and can communicate information to passengers through visual and/or auditory clues. In the early stages, some prototypes were developed to share situational awareness with AV researchers \cite{benderius2017best}. However for highly automated vehicles and considering people inside the cabin as mere passengers new iHMI has been developed with the goal of improving user experience, acceptance, and trust \cite{murali2022intelligent}. The iHMI has been in deep study to facilitate transferring from autonomous to manual mode and vice versa. But, to the best of our knowledge, there are no studies that have evaluated the effect of an iHMI when simultaneously interacting with other road users in a safety-critical scenario. In this work, we present results collected from real experimentation with an AV driving autonomously and interacting with a pedestrian in a crosswalk using simultaneously an iHMI and an eHMI. 

%% file: 3_description.tex
%\newpage
\section{Experiments description}

The goal of the study is to determine which factors including the internal and external HMIs and the behavior of the AV itself contribute to improve the level of confidence perceived by both pedestrians and passengers when interacting with an AV in a crosswalk area. With this goal in mind, a total of five experiments, four interaction experiments plus a control one were designed.

Our hypothesis is that the use of internal and external HMIs could help to increase the confidence of passengers and pedestrians when interacting with an AV. Furthermore, we believe that the AV's behavior plays a crucial role in instilling confidence. The smoother the behavior of the AV, the greater the confidence it imparts to both passengers and pedestrians.

The tests were designed in accordance with reproducibility standards, aiming to guarantee uniform interactions between the AV and all the participants. Following this criteria, the vehicle was programmed to change its speed profile at a specific point depending on the distance to the pedestrian, or more specifically, the distance to the edge of the crosswalk area. This mechanism enables the replication of a consistent behavior among all participants. The activation of the external and internal HMIs also relies on identical distance thresholds.

\begin{figure}[!t]
\centering
\includegraphics[width=\columnwidth]{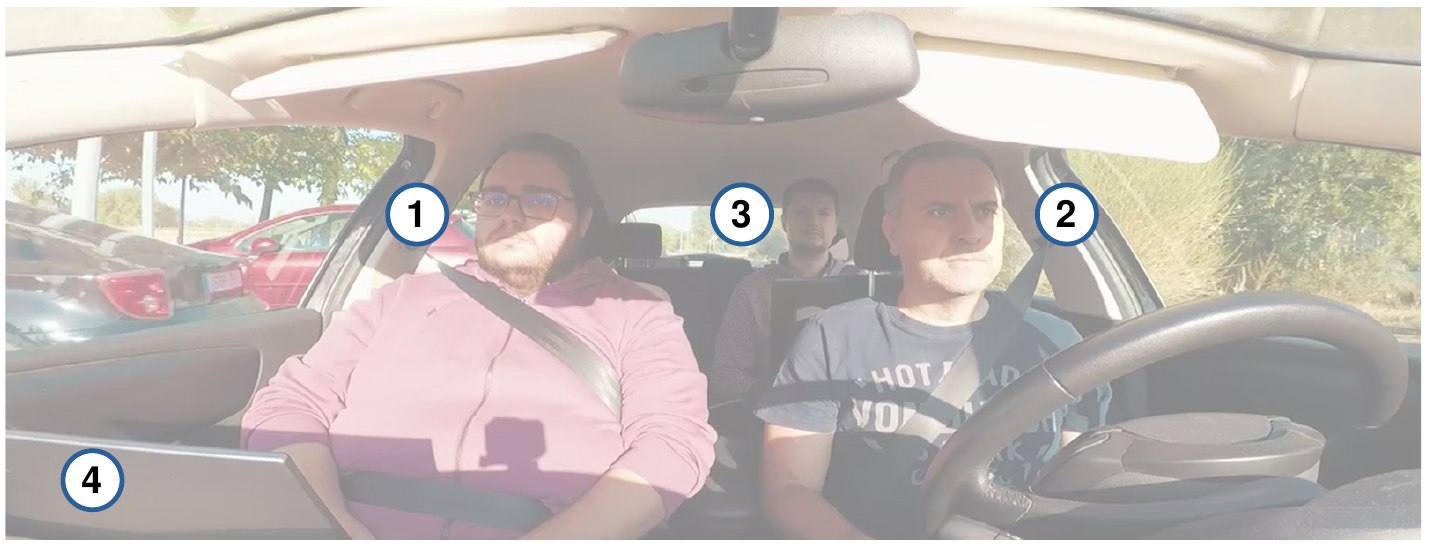}
\caption{View of the vehicle's passenger compartment. 1) The subject is seated in the passenger seat. 2) The backup driver is present but no action is required. 3) The system supervisor is seated in the rear. 4) Internal HMI.}
\label{fig:internal}
\end{figure}

The experiments were conducted using the autonomous and automated platform of the INVETT research group \cite{drivertive2018}, \cite{prevention2019}. This platform is a commercially available vehicle, modified to be externally controlled by a computer. It is equipped with a comprehensive setup for environmental detection and allows Real-Time Kinematic (RTK) positioning based on GPS \cite{prevention2019}. For the experiments conducted in this study, we used the front RGB camera and the GPS-based positioning system. Additionally, an internal camera mounted above the HMI was used to record the passengers' reactions. As shown in Fig. \ref{fig:internal}, the experiments were conducted with a backup driver for both legal and safety reasons. We also used a person seated in the rear seats to supervise the operation of the automated systems at all times. However, all subjects were duly informed that neither the backup driver nor the system supervisor were intervening during the vehicle's autonomous operation.

%The experiments have been carried out using the Autonomous platform of the INVETT research group. This vehicle is a commercial vehicle adapted to be externally controlled by a computer. It has a complete setup for the detection of the environment and allows Real-Time Kinematic (RTK) positioning based on GPS. The front RGB camera and the GPS-based positioning system have been used for the experiments conducted in this study. Internally, a camera mounted above the HMI recorded the passengers' reactions.

%These scenarios aim to measure the perceived confidence and consequently the level of trust in the AV and ultimately the acceptance of this type of new transport system.
%These scenarios aim to measure the acceptance, trust and perceived safety or risk of passengers and pedestrians interacting with the vehicle by using different types of explicit and implicit communication mechanisms.

\subsection{Use case scenario}
\begin{figure}[!t]
\centering
\includegraphics[width=\columnwidth]{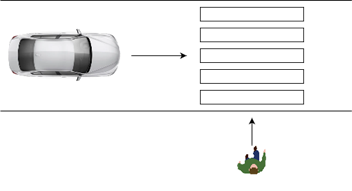}
\caption{Schematic representation of the experiment use case.}
\label{fig:usecase1}
\end{figure}

The use case evaluated in this study is a complete stop at a crosswalk yielding to a pedestrian that approaches, stands, or crosses the crosswalk. The vehicle drives at a constant speed and at a specific point, (depending on the experiment) reduces its speed to finally stop before the crosswalk, even if the pedestrian chooses not to cross and stands at the limit of the sidewalk. The fact that the vehicle is going to stop under any circumstance is deliberately omitted to the subjects of the experiment to preserve the perception of risk. Fig. \ref{fig:usecase1} shows a schematic representation of the interaction between the AV, the passenger, and the pedestrian.

\begin{figure}[!t]
\centering
\includegraphics[width=\columnwidth]{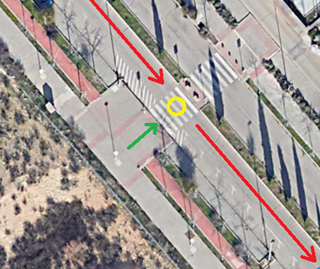}
\caption{Location of the experimentation area (40°30'58.1"N 3°20'40.6"W). The red arrow represents the traveling direction of the automated vehicle, the green arrow the path of the pedestrian, and the yellow circle the interaction area over the crosswalk.}
\label{fig:location}
\end{figure}

The ground test site must meet certain requirements. First, the pedestrian must not be influenced by any other vehicles. Therefore, a single-lane road is necessary. There must be a crosswalk to perform the tests for the use case. A low-traffic area is also desired so as not to block the road during the trials and not to have other vehicles queuing. Based on these requirements, tests were carried out in the vicinity of the Polytechnic School within the Technological Campus of the University. Figure \ref{fig:location} shows the designated area. The red arrow shows the trajectory of the AV, the green arrow the pedestrian's path, and the yellow circle marks the crosswalk area. While Google Maps aerial images indicate an empty parking area, it was, in fact, occupied by cars during the test. The crosswalk in question is linked to another one where potential interactions with vehicles moving in the opposite direction may occur. Participants were instructed to only cross to the central island and avoid proceeding further to remove undesired interactions and maintain the pedestrian focus on the experiment. 

\subsection{Vehicle communication setup}
The vehicle is equipped with two HMIs to interact with the passenger inside the AV and the road users. The external HMI (or eHMI) is called GRAIL\cite{drivertive2018}. It is a directionable RGB LED strip located in the front bumper of the vehicle to interact and communicate with the road users. The internal HMI (or iHMI) consists of a 16-inch audio-capable screen located on the dashboard in front of the co-pilot to interact with the passenger. Both the eHMI and the iHMI devices are explicit communication tools. In addition, vehicle dynamics are considered as an implicit communication tool and are consequently explored.

\subsubsection{External HMI (eHMI)}
\begin{figure}[!t]
\centering
\includegraphics[width=\columnwidth]{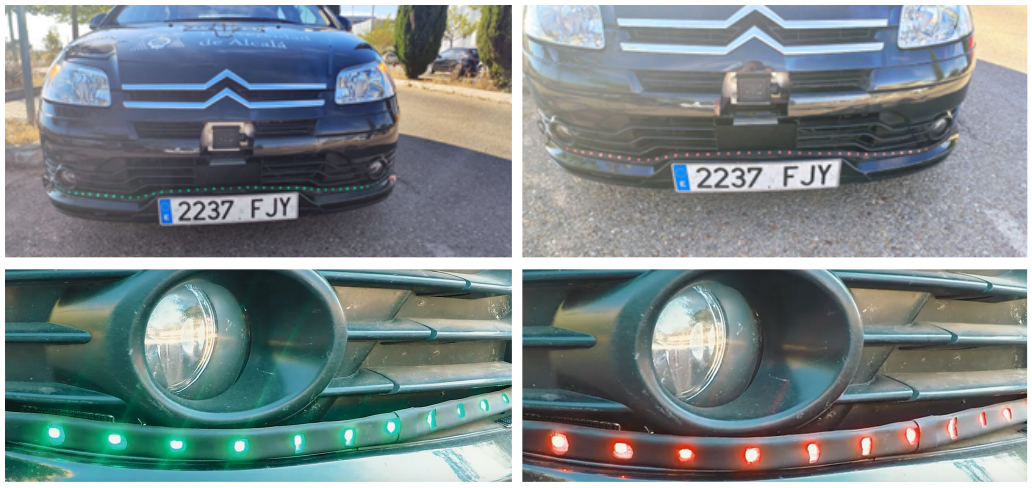}
\caption{External HMI (eHMI) - Left \textit{solid green} state and right \textit{solid red} state.}
\label{fig:GRAIL}
\end{figure}

The external communication device (GRAIL \cite{drivertive2018}) was configured with three possible states; \textit{off}, \textit{solid red}, and \textit{solid green}. When the state is \textit{off} the LED strip looks like a black strip on the black bumper of the vehicle and it is practically not visible. When GRAIL is actively used, the strip emits a solid red or green light. The \textit{solid red} state is used while the vehicle is traveling at its cruising speed. The \textit{solid green} state is used when the vehicle changes its behavior and starts to slow down. Note that the goal of the eHMI is not to establish a target-based communication with the pedestrian, but to convey the vehicle´s intentions. Fig. \ref{fig:GRAIL} shows the two active states of the GRAIL device. 
The sequence of the eHMI state when it is used in the experiments is: \textit{off} $\rightarrow$ \textit{solid red} $\rightarrow$ \textit{solid green} $\rightarrow$ \textit{off}. 

% \subsubsection{Explicit communication}
% The external communication interface (GRAIL) \cite{GRAIL} has three states, off, solid red, and solid green. When off it looks like a black strip on the black front bumper of the vehicle. When GRAIL is actively used, the strip emits a solid red light when moving at its cruising speed. When the "event" is activated, it turns green, indicating that the vehicle is changing its behavior and is slowing down. See examples of GRAIL in Figure \ref{fig:GRAIL}.

\subsubsection{Internal HMI (iHMI)}
\begin{figure}[!t]
\centering
\includegraphics[width=\columnwidth]{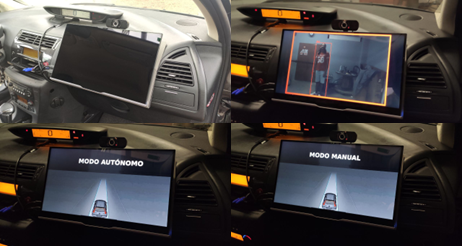}
\caption{Internal HMI (iHMI) states: top left \textit{off}, top right \textit{pedestrian detected}, bottom left \textit{autonomous mode}, and bottom right \textit{manual mode}.}
\label{fig:internalHMI}
\end{figure}

The internal communication device is a 16-inch audio-capable screen located in front of the co-pilot over the dashboard. It has four possible states; \textit{off}, \textit{autonomous mode}, \textit{manual mode}, and \textit{pedestrian detected}. Fig. \ref{fig:internalHMI} shows the four possible states of the iHMI. The default state is \textit{off}, and the screen remains black with no sounds. When it is actively used the screen shows different images or video sources together with audio messages. The \textit{autonomous mode} plays the sentence \textit{"autonomous mode activated"} once at the time the screen changes to its corresponding static image showing the text \textit{AUTONOMOUS MODE} in the Spanish language. When the state changes to \textit{manual mode} the sentence \textit{"autonomous mode deactivated"} is played at the time the screen changes to its corresponding static image showing the text \textit{MANUAL MODE} also in the Spanish language. The state \textit{pedestrian detected} is triggered at a specific distance based on the experiment requirements playing the sentence \textit{"pedestrian detected"} while the exterior camera video stream is reproduced on the iHMI together with a red bounding box over the detected pedestrian and a flashing red rectangle around the limit of the screen.
The sequence of the iHMI state when it is used in the experiments is: \textit{off} $\rightarrow$ \textit{autonomous mode} $\rightarrow$ \textit{pedestrian detected} $\rightarrow$ \textit{manual mode}.

%of the iHMIs is \textit{off} the screen remains black with no sounds. When it is actively used the \textit{} it shows different images or video sources together with audio messages.  when the vehicle starts to move and reproduces an audio message saying: “Autonomous mode activated”. When the “event” is triggered, the HMI shows a live video stream of the scene in front of the vehicle with the detection of the pedestrian depicted with a red rectangle, and the following audio is played: “pedestrian detected”. %The “event” refers to the position point where the vehicle starts to brake.

\subsubsection{Vehicle Dynamics}
Vehicle dynamics can be used as an implicit way of communication. In these experiments, the message to be communicated is the intention of the vehicle to stop (or not) at the crosswalk and to yield to the pedestrian. Two alternatives have been proposed to explore this kind of communication.

The gentle braking maneuver. This braking maneuver is characterized by a smooth and early deceleration. This situation replicates the performance of early detection systems that can provide sufficient anticipation by detecting and predicting the intention of the pedestrian. Consequently, the anticipation of the breaking maneuver leads to increased comfort and safety for both passengers and road users.

The aggressive braking maneuver. It is characterized by a delayed and stronger deceleration, in opposition to the early braking maneuver. This situation replicated the performance of classic Advance Driver Assistance Systems (ADAS) or last-second reaction systems. The delayed initiation of the braking maneuver causes a stronger deceleration to stop the vehicle at the limit of the crosswalk compared with the gentle braking maneuver.

For practical purposes, these two braking maneuvers have been generated following a constant acceleration (deceleration) movement according to the desired distance to the stop point at the limit of the crosswalk area. The trigger distances to the stop point are 40 meters for the gentle braking maneuver and 20 meters for the aggressive one. The vehicle travels at 30 $km/h$ before the initiation of the braking maneuver. Consequently, the constant acceleration is -0.86 $m\cdot s^{-2}$ and -1.73 $m\cdot s^{-2}$ for the gentle and aggressive braking maneuvers, respectively.

\begin{table}[b]
\renewcommand{\arraystretch}{1.5}
\caption{Configuration of Experimentation Tests}
\begin{center}
\begin{tabular}{ccccc}
%\hline
\textbf{Test}& \textbf{Braking} & \multicolumn{2}{c}{\textbf{Explicit}} & \textbf{Stop}\\
%\cline{3-4} 
\textbf{Number} & \textbf{Maneuver} & \textbf{Internal}& \textbf{External} & \\
\hline
0   & -             & -     & -     & No \\
\hline
1   & Gentle        & -     & -     & Yes\\
2   & Aggressive    & -     & -     & Yes\\
\hline
3   & Gentle        & HMI   & GRAIL & Yes\\
4   & Aggressive    & HMI   & GRAIL & Yes\\
%\hline
\end{tabular}
\label{tab:testsetup}
\end{center}
\end{table}

\subsection{Test configuration}
Following the definition of the use case and the possibilities to use the explicit and implicit ways of communication, several tests have been proposed to evaluate how each of these features affects the passenger's and pedestrian's experience interacting with the AV. Note that the iHMI and the eHMI are independent devices that produce independent effects on the passenger and the pedestrian, respectively. The passenger does not perceive the eHMI and the pedestrian does not perceive the iHMI, and more importantly, none of them has the ability to affect the behavior of the other. For this reason, the combination of the three sources of variability is reduced to two, resulting in a total combination of four experiments. Table \ref{tab:testsetup} summarizes the configuration for each test. In addition to these four experiments, a preliminary test denoted by test 0 was added to create the illusion that the vehicle could cross through the crosswalk without stopping or yielding to the pedestrian.

% \begin{table}[!t]
% \caption{Contribution to increase of pedestrian´s perceived safety\label{tab:testconfig}}
% \centering
% \begin{tabular}{r|cccc|c}
% Test   & \multicolumn{2}{c}{Interior}   & \multicolumn{2}{c|}{Exterior} & Stop\\
%        & Implicit       & Explicit      & Implicit       & Explicit     &     \\
% \hline
% 0           & None      & None          & None      & None  & No \\
% 1           & Gentle    & None          & Gentle    & None  & Yes \\
% 2           & Aggressive& None          & Aggressive& None  & Yes \\
% 3           & Gentle    & Audio/Visual  & Gentle    & GRAIL & Yes \\
% 4           & Aggressive& Audio/Visual  & Aggressive& GRAIL & Yes \\
% \end{tabular}
% \end{table}

Tests from 1 to 4 were performed in random order. Test 0 was always performed first. The experimental subjects do not know the order or configuration of each test with the exception of test 0.

\subsection{Participants}  %% REVIEW HERE
Participants were recruited from university staff, friends, relatives, and others. They must be over 18 years of age. They were informed of the purpose of the study and what was expected to occur during the study. To formally comply with legal requirements, an informed consent and an informed consent statement were developed to record evidence of the user acceptance and to anonymize the participant's personal information by assigning an anonymous ID.

Participants were instructed to participate in couples and to play both pedestrian and passenger roles. Firstly, one of them performs the passenger role while the other performs the pedestrian role. After finishing the complete set of tests, the participants swap roles to perform the complete set of tests again in the same random order. With this mechanism, we can observe differences in the perception of the interaction between those who were first passengers or pedestrians in case those differences exist.

A total of 34 people joined the experiment but two of them could not complete the whole set of tests due to technical problems and their information was discarded. The number of subjects is $N=32$ (18 men and 14 women) with an age distribution $\mu=39.7$ and $\sigma=12.6$ years.

\subsection{Briefing}
Participants were given an explanation of what to expect and what to do in the experiments. This information was repeatedly without variation to all the subjects with the goal of not introducing any external source of change in the experimentation. The participants received an explanation for the tests conducted as a passenger and another as a pedestrian.

As passengers, they were told:
\begin{enumerate}
\item There is an HMI which consists of a screen that can display images and reproduce messages.
\item There is a webcam recording the co-pilot seat area.
\item There is a backup driver just to comply with legal requirements.
\item The backup driver is instructed not to intervene unless critical and imminent damage.
\item The vehicle will drive itself autonomously and interact with the pedestrian.
\end{enumerate}

As pedestrians, they were told:
\begin{enumerate}
\item There is an HMI consisting of an LED strip that could be off, red, or green (all three modes are displayed to the pedestrian prior to testing).
\item If the LED is off there is no information about the behavior of the vehicle. If the LED is red, it means that the vehicle is driving at its cruising speed. If the LED is green, it means that the vehicle has detected something in its path and is acting accordingly (note that the specific behavior of the vehicle is not stated).
\item There is a camera on the vehicle that can see you and record you.
\item There is a backup driver just to comply with legal requirements.
\item The backup driver is instructed not to intervene unless critical and imminent damage.
\item The vehicle will drive itself autonomously and interact with the pedestrian.
\end{enumerate}

Three staff members and two participants are required to conduct the experiment. The participants interact with the vehicle as a passenger and as a pedestrian and the staff is responsible for (1) backup driver, (2) commanding the tests in the AV software, and (3) letting the pedestrian know when to start moving into the interaction area. The pedestrian stands on the sidewalk backward to the crosswalk with no information about the traffic status. At a specific position of the AV the pedestrian is requested to turn around and walk towards the crosswalk area generating a proper and credible interaction.

\section{Experiment evaluation}
The experiments were evaluated using two different sources of information. Questionnaires are one of the sources of data used for the analysis. With these elements, the analysis was made using subjective information about the interaction from the participant's point of view. 
Direct measures recorded from the AV's sensors are also used to complete the data for the analysis. This information is objective and allows us to objectively analyze the interactions.

\subsection{Questionnaires}
Two questionnaires were developed to record the participants' opinions. The first questionnaire records general knowledge about AVs, past experiences and interactions with AVs, and expectations. This questionnaire is filled out by participants before and after the experimentation. The goal is to verify with a manipulation check if the participants correctly understood that they have interacted with an AV and to evaluate how their experiences and expectations about AVs have changed after the experimentation. 

The second questionnaire is designed to assess passenger and pedestrian confidence and feelings about the interaction with the AV after each test. These questions were formulated using the 7-step Likert scale when possible. Right after each test and before starting the following one all the questions were answered. This questionnaire has three questions that are answered when interacting as a pedestrian and four for passenger interaction (see Appendices \ref{appendix:questions1} and \ref{appendix:questions2}).

\subsection{Direct measuring}
Different sources of information are needed to directly measure the interaction between the AV and the participants in addition to the questionnaires. For each experiment, the following information was recorded by the AV software:
\begin{itemize}
    \item AV logging file including vehicle position, speed, and distance to the pedestrian.
    \item In-vehicle external video and time logging.
    \item Internal video of the co-pilot area and time logging.
    \item Communication log between AV and HMI systems.
\end{itemize}

\begin{figure}
\centerline{\includegraphics[width=\columnwidth]{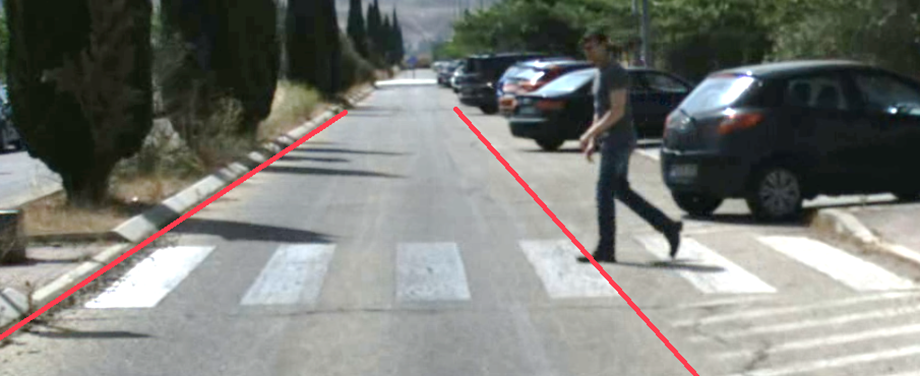}}
\caption{Crossing event example. Vehicle lane is defined by road marks unequivocally for all the experiments.}
\label{fig:vehicle_lane}
\end{figure}

By processing video and data information it is possible to determine when and where the pedestrian decides to cross through the crosswalk. This event is of utmost importance because it is ultimately affected by the type of interaction between the AV and the pedestrian and can reveal how the eHMI and the implicit communication affect the interaction.

The crossing decision event is defined as the moment in which the pedestrian makes the mental decision to cross. We follow the hypothesis that the decision to cross is a hidden state with an external and delayed manifestation that can be observed. The delay between the decision and its external manifestation can vary depending on the person and the situation. Alternatively, to the crossing decision event, we propose to use the crossing event as the metric to evaluate the behavior of the pedestrian using direct measurements in the experiments. The crossing event is defined as the frame in which the pedestrian enters the vehicle lane and physically exposes his/her body to a potential and real injury. The background idea is that an early crossing decision will produce an early crossing event and a late crossing decision will produce a late crossing event. The main difference is that the crossing event is not a hidden state. It is directly observable and can be unequivocally identified when the vehicle lane is defined using the road marks. Figure \ref{fig:vehicle_lane} shows the vehicle lane boundary at the crossing event frame. It is defined as the moment the pedestrian enters the area of the vehicle’s lane. See figure \ref{fig:example_interacion_exterior} for a crossing sequence description example.

Several physical variables can be directly used or computed to represent the interaction such as the distance between the AV and the pedestrian, the approaching speed which is equal to the vehicle speed as far as the pedestrian travels perpendicularly to the vehicle trajectory, and a combination of them which is the so-called Time To Collision (TTC). The TTC is a magnitude measured in seconds that represents how many seconds the vehicle needs to hit the pedestrian if the vehicle continues at the same speed. Its calculation is simple and it is the quotient of the distance to the pedestrian $d$ over the vehicle speed $v$.
\begin{equation}
TTC=d/v
\end{equation}

TTC is a \textit{vehicle-centric} variable that depends on the vehicle's speed and the distance to the pedestrian. It can effectively measure the potential risk perceived by the pedestrian. However, if we analyze the interaction from the pedestrian point of view, and more specifically from the optical point of view, the volume of the vehicle (or its solid angle) must be included to correctly measure the potential risk perceived by the pedestrian.

\begin{figure}[!t]
\centering
\includegraphics[width=\columnwidth]{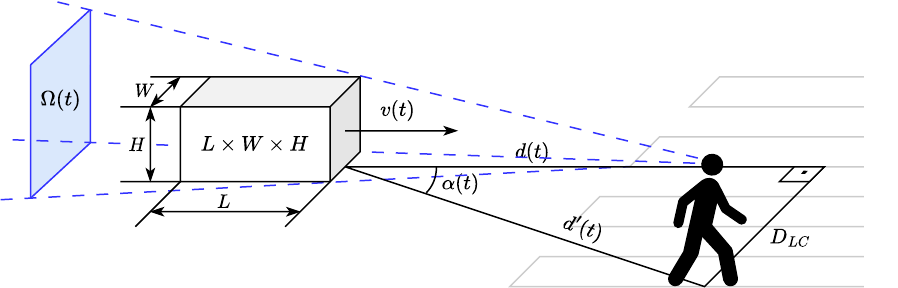}
\caption{Schematic representation of a simplified moving vehicle and its observation from the pedestrian's point of view.}
\label{fig_2}
\end{figure}

Figure \ref{fig_2} shows a representation of a vehicle simplified by a rectangular parallelepiped with dimensions $L\times W\times H$ (Height, Width, and Length) driving towards a crosswalk area at a distance $d(t)$ with a given velocity $v(t)$. The constant $D_{LC}$ represents the distance from the pedestrian standing point to the vehicle's lane center. 
The pedestrian observation angle $\alpha(t)$, formed by the vehicle's moving direction, and the pedestrian observation line is computed according to eq. \ref{eq:scenario1}.
\begin{equation}
\label{eq:scenario1}
    \alpha(t) = \tan^{-1}{D_{LC}/d(t)}
\end{equation}
Given the observation angle $\alpha(t)$, the apparent distance $d'(t)$ between the vehicle and the pedestrian can be computed as it is shown in eq. \ref{eq:scenario2}.
\begin{equation}
\label{eq:scenario2}
    d'(t) = d(t) / {\cos{\alpha(t)}}
\end{equation}
$\Omega(t)$ is the solid angle represented by the vehicle being observed from the pedestrian's point of view and it is computed as:
\begin{equation}
%    \Omega(t) = H \cos{\alpha(t)}\dfrac{W\cos{\alpha(t)} + L\sin(\alpha(t))}{d'(t)}
    \Omega(t) =\dfrac{A(t)}{r(t)^2} = \dfrac{H\left( W\cos{\alpha(t)} + L\sin\alpha(t) \right)}{d'(t)^2}
\end{equation}
The change rate of the solid angle, $d\Omega(t)/dt$, is often used as a parameter to measure the reaction to a moving object. Usually, 0.2 $rd/s$ is considered the threshold to visually trigger a reaction.

\subsection{Experiment samples}
\begin{figure}
\centering
\includegraphics[width=\columnwidth]{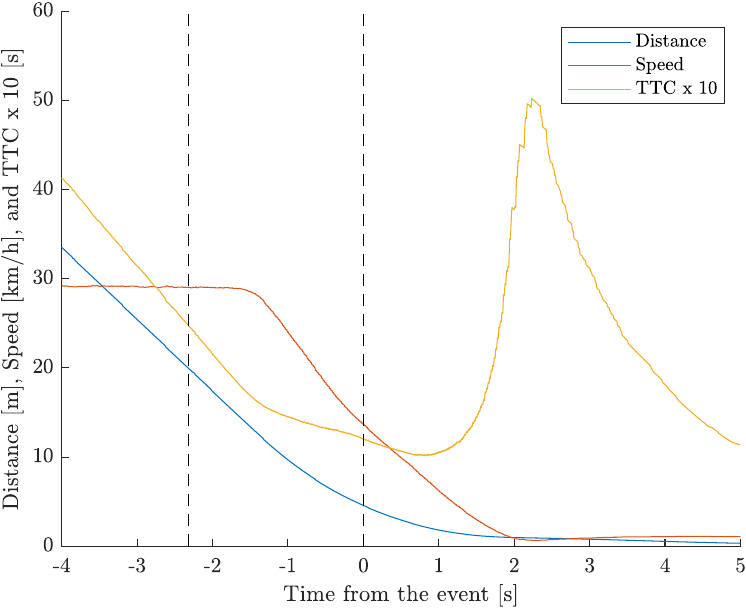}
\caption{Observed Time To Collision based on the distance to the pedestrian and the vehicle speed for one of the experiments.}
\label{fig:example_ttc}
\end{figure}

\begin{figure}
\centering
\includegraphics[width=\columnwidth]{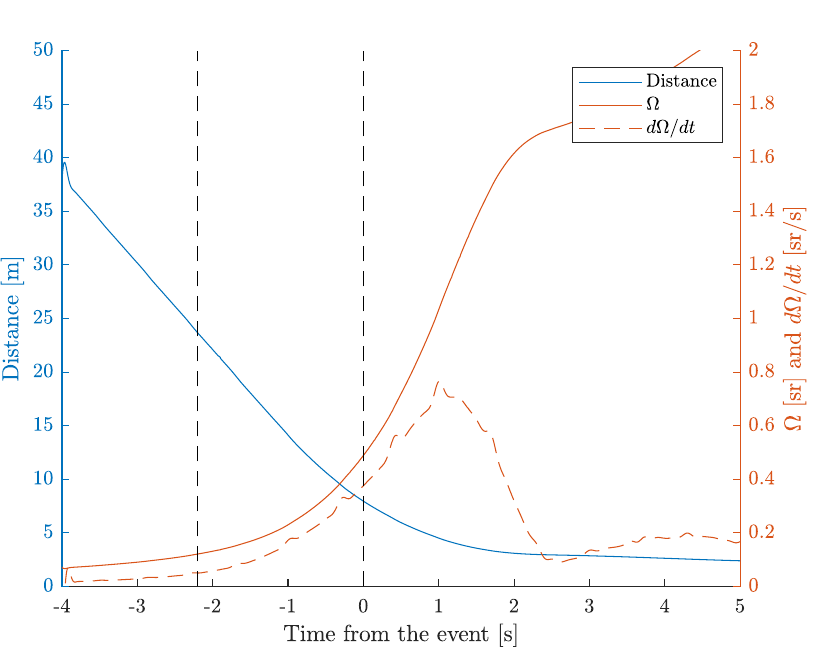}
\caption{Observed vehicle solid angle and its temporal change rate calculated based on the distance to the pedestrian for one of the experiments.}
\label{fig:example_angle}
\end{figure}

\begin{figure*}[!t]
\centering
\subfloat[]{\includegraphics[width=0.22\textwidth]{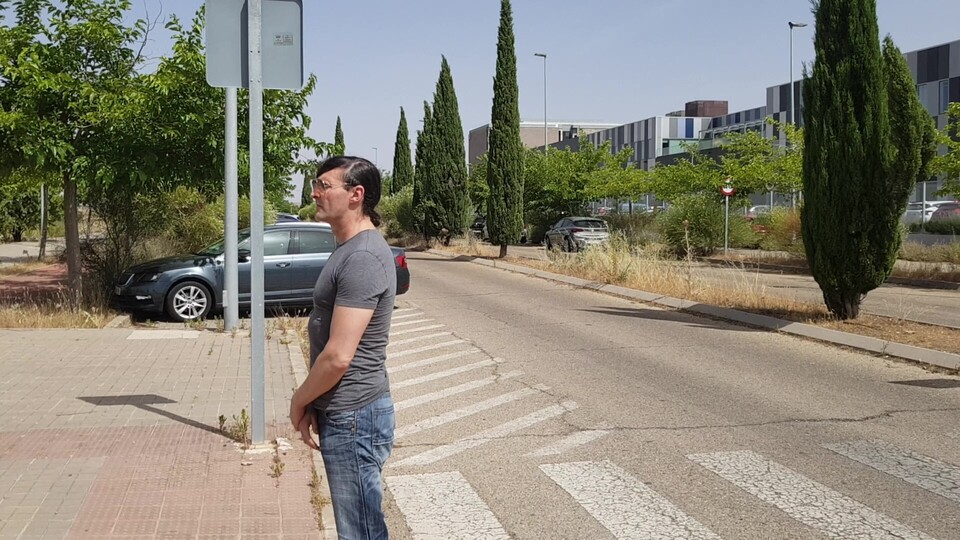}%
\label{fig_first_casea}}
\hfil
\subfloat[]{\includegraphics[width=0.22\textwidth]{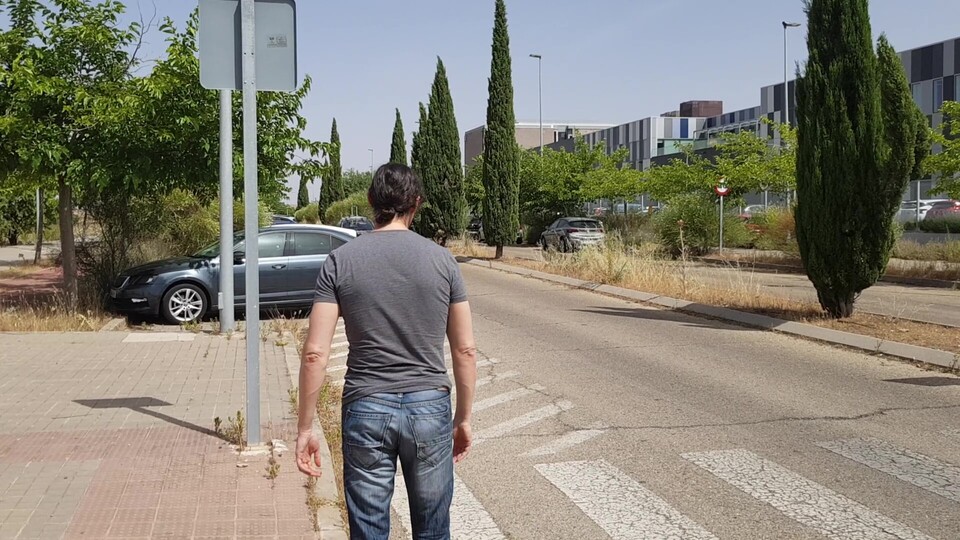}%
\label{fig_first_caseb}}
\hfil
\subfloat[]{\includegraphics[width=0.22\textwidth]{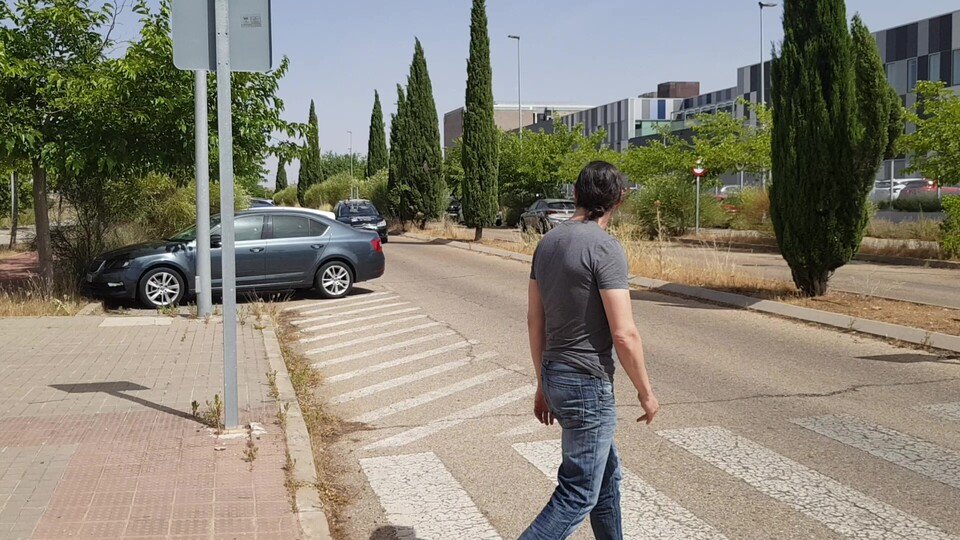}%
\label{fig_first_casec}}
\hfil
\subfloat[]{\includegraphics[width=0.22\textwidth]{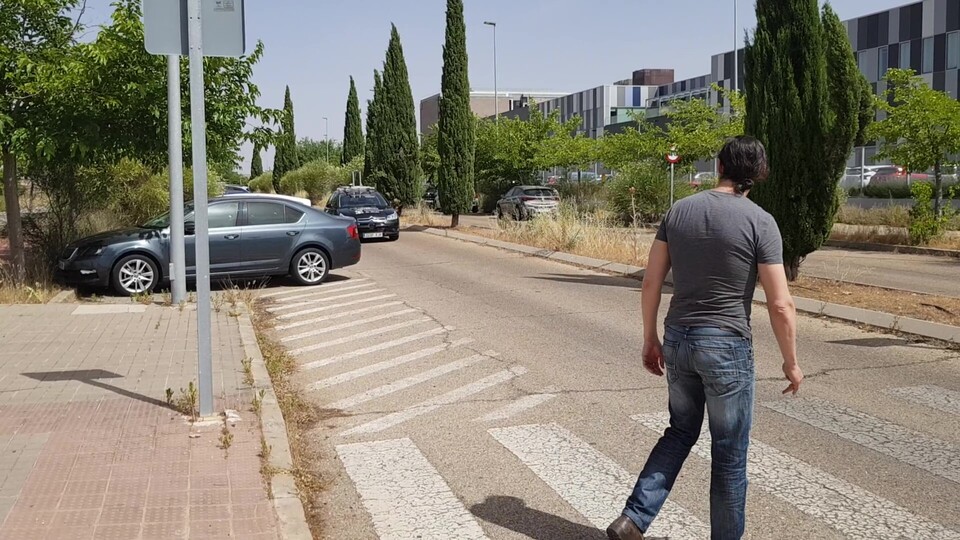}%
\label{fig_first_cased}}
\hfil
\subfloat[]{\includegraphics[width=0.22\textwidth]{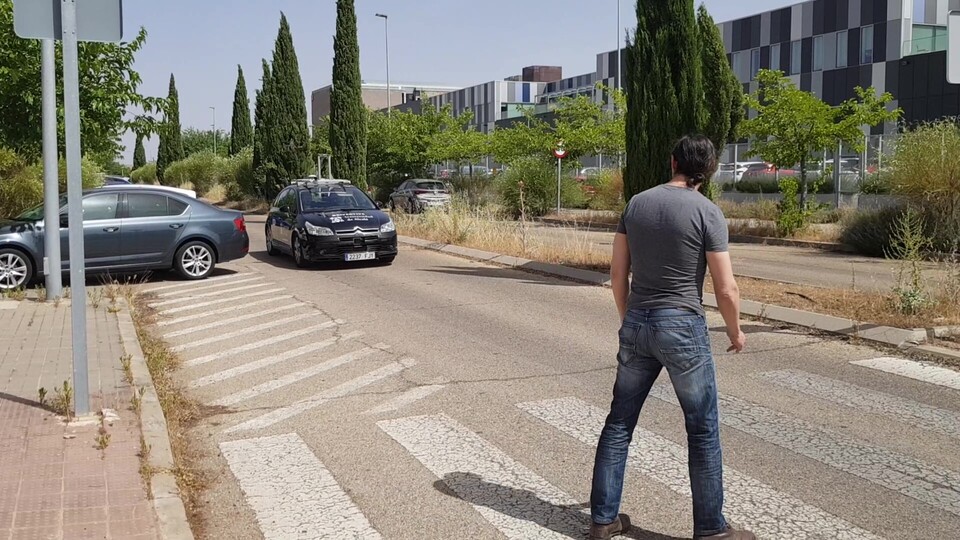}%
\label{fig_first_casee}}
\hfil
\subfloat[]{\includegraphics[width=0.22\textwidth]{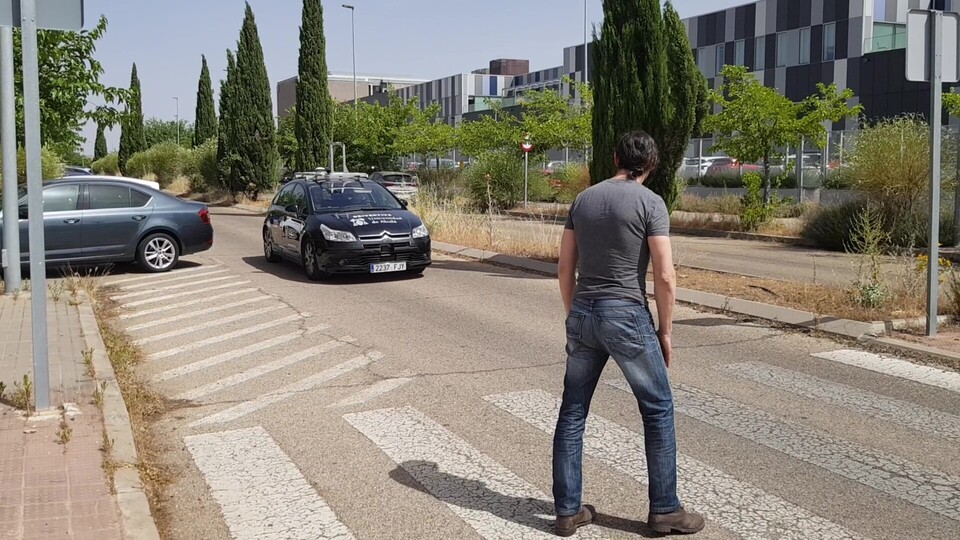}%
\label{fig_first_casef}}
\hfil
\subfloat[]{\includegraphics[width=0.22\textwidth]{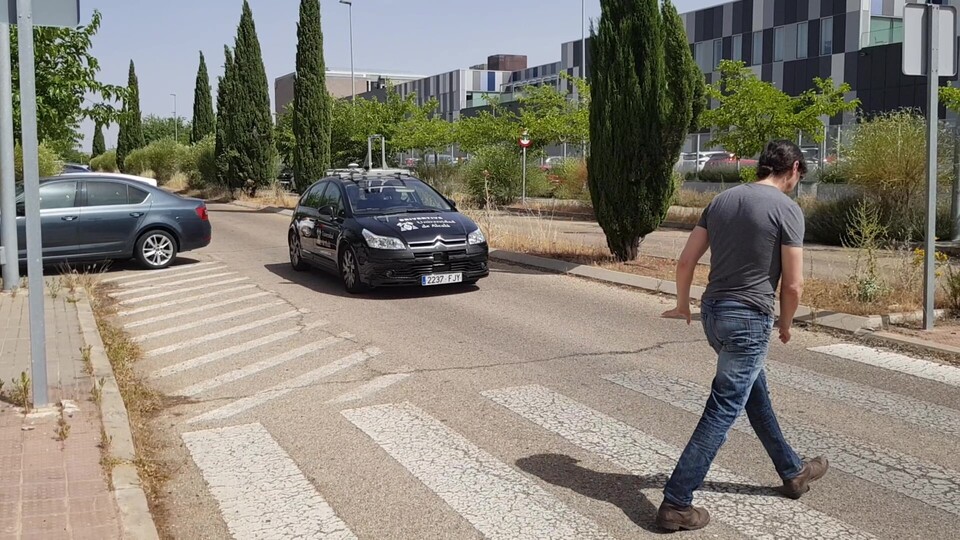}%
\label{fig_first_caseg}}
\hfil
\subfloat[]{\includegraphics[width=0.22\textwidth]{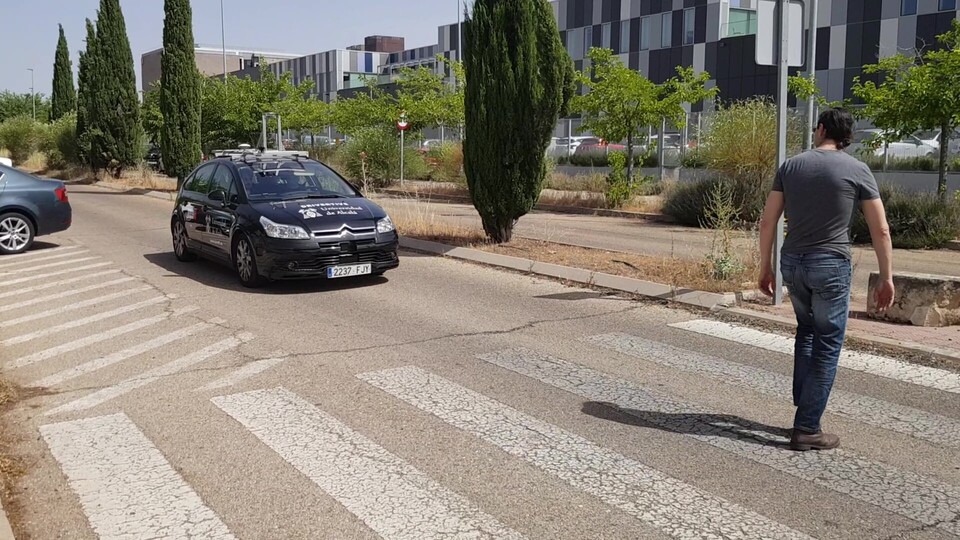}%
\label{fig_first_caseh}}
\caption{Example of vehicle-pedestrian interaction - exterior camera. (a) Initial position of pedestrian back to the crosswalk. (b) The pedestrian turns and faces the crosswalk. (c) The pedestrian starts walking and sees the vehicle approaching. (d) At this point, the pedestrian hesitates to cross. (e) The pedestrian is still waiting for the vehicle's reaction. (f) The pedestrian does not feel comfortable crossing while the vehicle is moving. (g) The Pedestrian decides to cross when the vehicle is almost stopped. (h) The Pedestrian crosses the crosswalk.}
\label{fig:example_interacion_exterior}
\end{figure*}

\begin{figure*}[!t]
\centering
\subfloat[]{\includegraphics[width=0.22\textwidth]{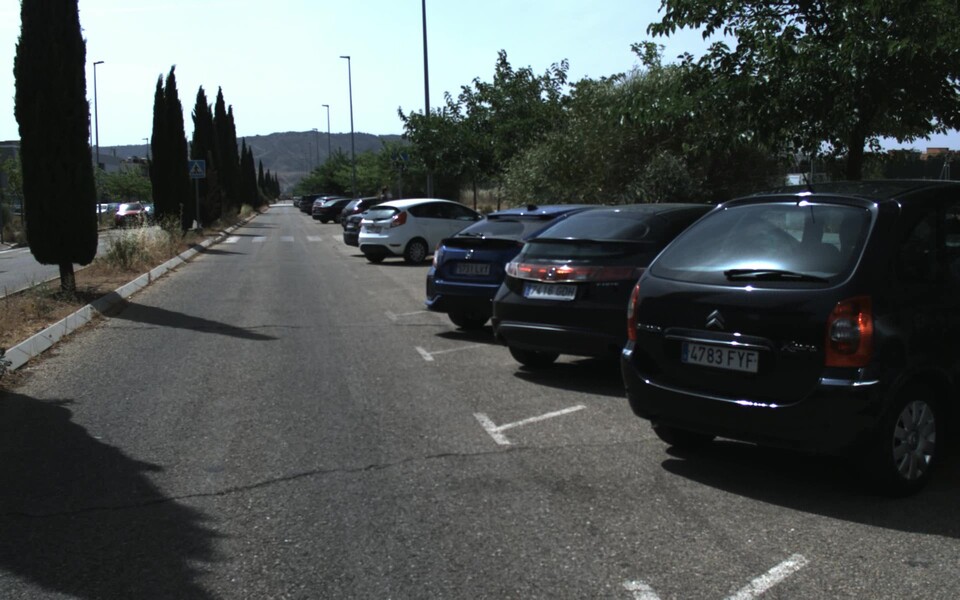}%
\label{fig_second_casea}}
\hfil
\subfloat[]{\includegraphics[width=0.22\textwidth]{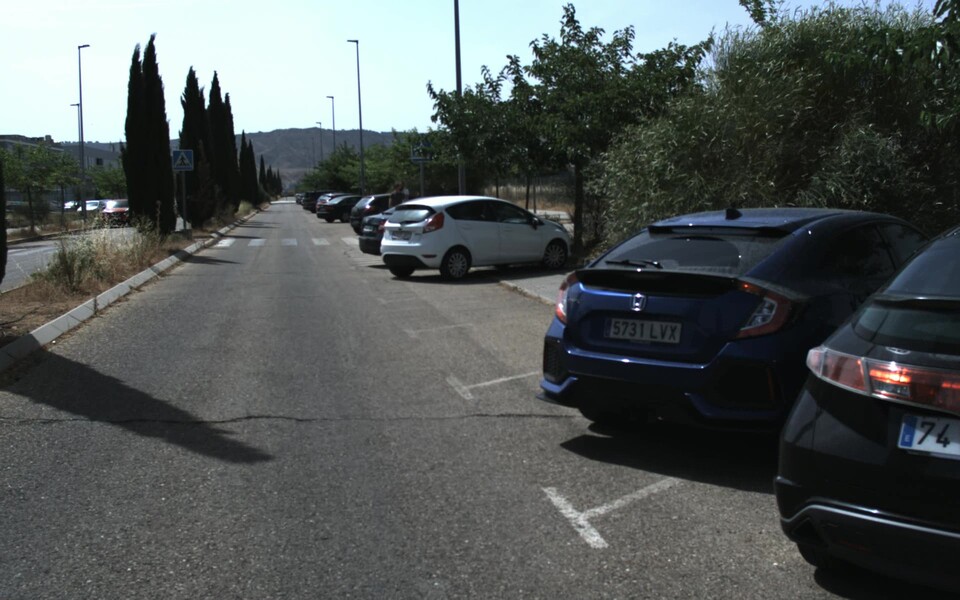}%
\label{fig_second_caseb}}
\hfil
\subfloat[]{\includegraphics[width=0.22\textwidth]{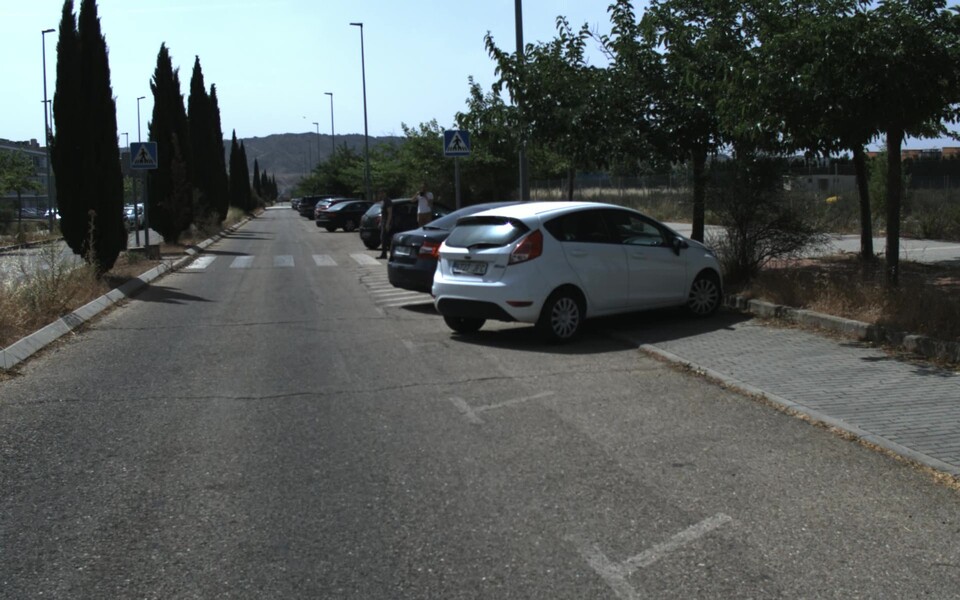}%
\label{fig_second_casec}}
\hfil
\subfloat[]{\includegraphics[width=0.22\textwidth]{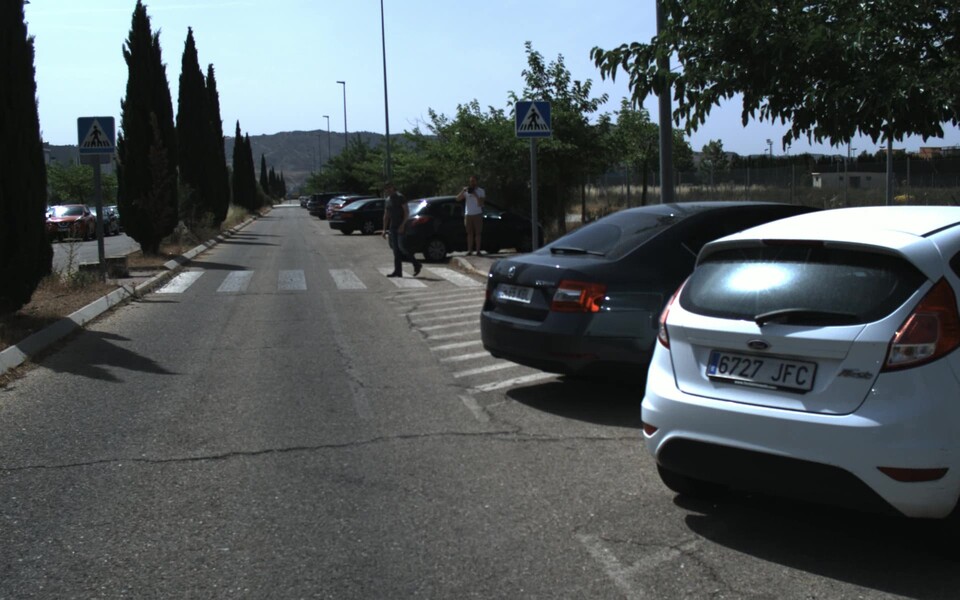}%
\label{fig_second_cased}}
\hfil
\subfloat[]{\includegraphics[width=0.22\textwidth]{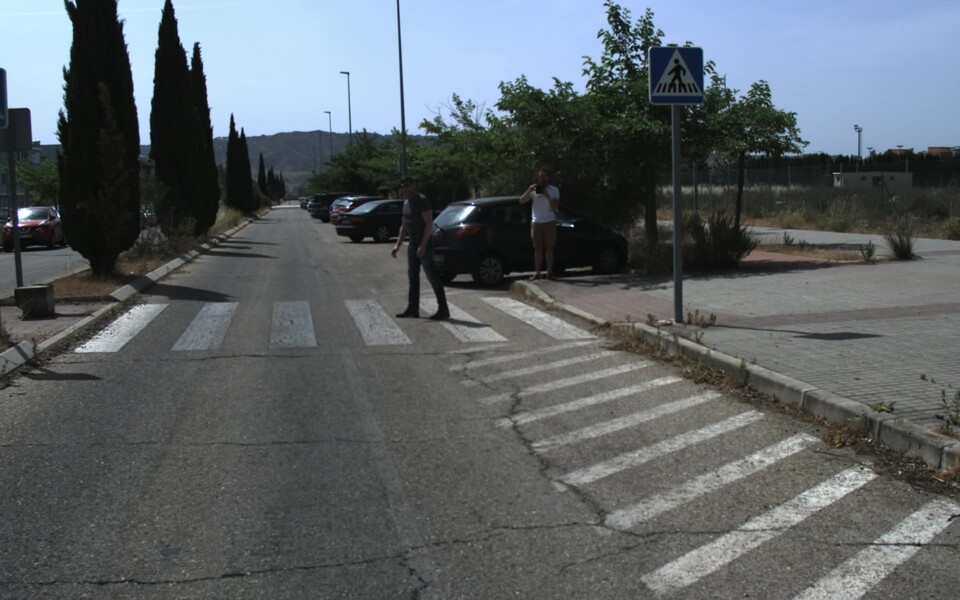}%
\label{fig_second_casee}}
\hfil
\subfloat[]{\includegraphics[width=0.22\textwidth]{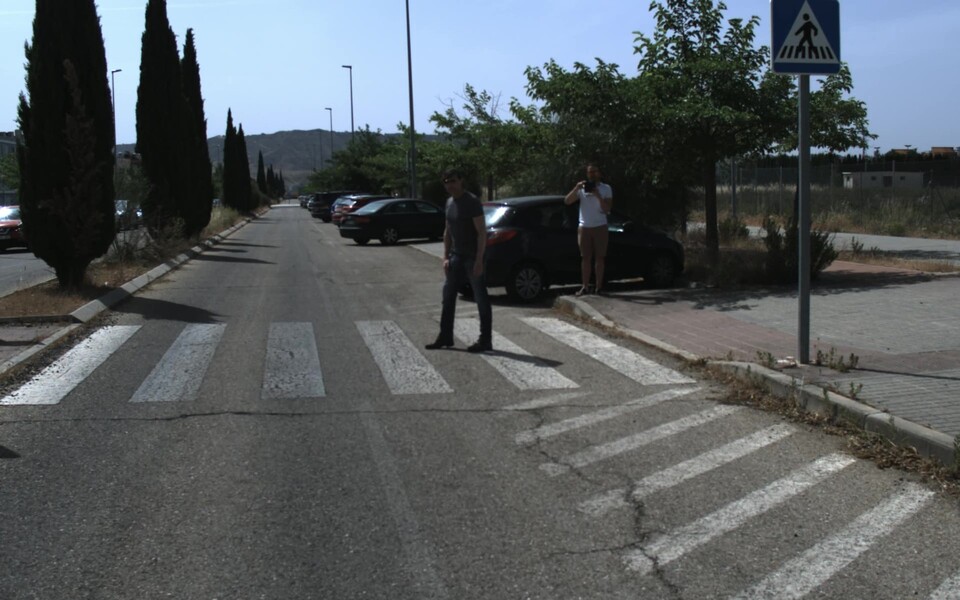}%
\label{fig_second_casef}}
\hfil
\subfloat[]{\includegraphics[width=0.22\textwidth]{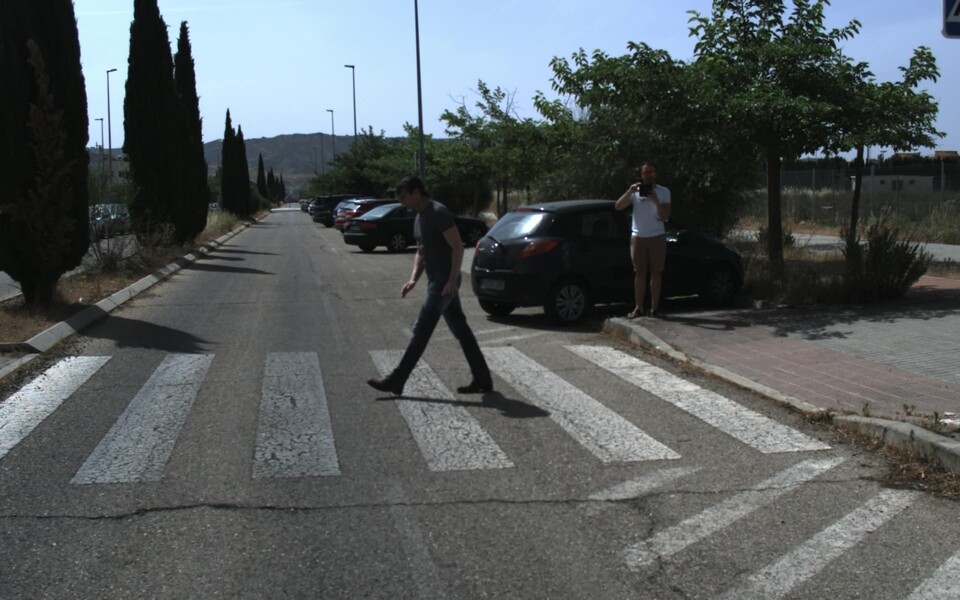}%
\label{fig_second_caseg}}
\hfil
\subfloat[]{\includegraphics[width=0.22\textwidth]{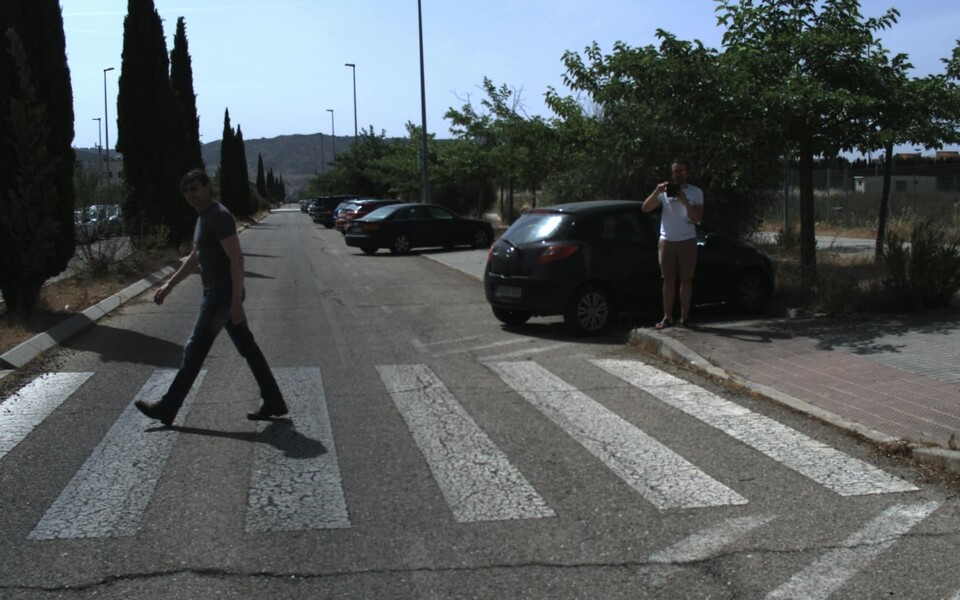}%
\label{fig_second_caseh}}
\caption{Example of vehicle-pedestrian interaction - in-vehicle camera. Images from (a) to (h) correspond to the same frames and descriptions as in figure \ref{fig:example_interacion_exterior}.}
\label{fig:example_interaction_invehicle}
\end{figure*}

\begin{figure*}[!t]
\centering
\subfloat[]{\includegraphics[width=0.251\textwidth]{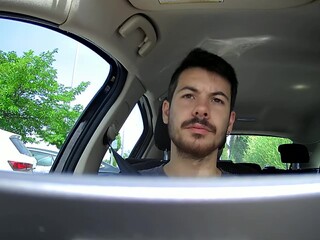}%
\label{fig_pas_a}}
\hfil
\subfloat[]{\includegraphics[width=0.251\textwidth]{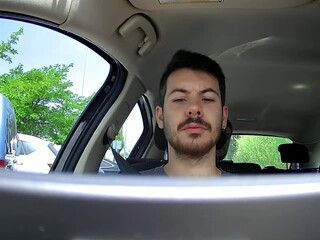}%
\label{fig_pas_b}}
\hfil
\subfloat[]{\includegraphics[width=0.251\textwidth]{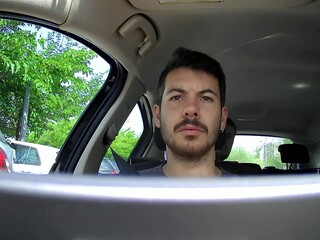}%
\label{fig_pas_c}}
\hfil
\subfloat[]{\includegraphics[width=0.251\textwidth]{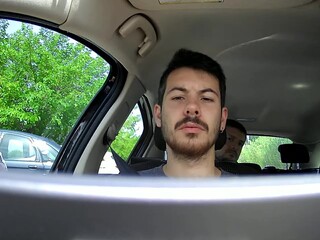}%
\label{fig_pas_d}}
\hfil
\subfloat[]{\includegraphics[width=0.251\textwidth]{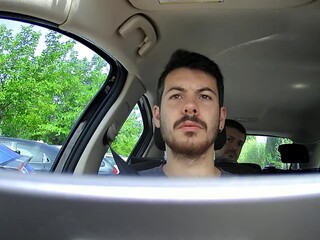}%
\label{fig_pas_e}}
\hfil
\subfloat[]{\includegraphics[width=0.251\textwidth]{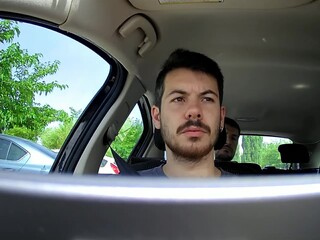}%
\label{fig_pas_f}}
\caption{Example of vehicle-passenger interaction - Internal video of the co-pilot area. (a) The passenger, just before the start of the test, looks forward. (b) “Autonomous mode on” displayed on the screen and played back on the speakers. (c) The test begins and the vehicle starts to move forward. (d) Outdoor video with detections displayed on the iHMI and "Pedestrian detected" played on speakers. (e) The passenger looks at the pedestrian after the interaction with the iHMI. (f) The passenger follows the pedestrian with his eyes as s/he crosses the road.}
\label{fig:example_interaction_passenger}
\end{figure*}

Figs. \ref{fig:example_interacion_exterior}, \ref{fig:example_interaction_invehicle}, and \ref{fig:example_interaction_passenger} show different instances of one of the test from three different perspectives. Figs. \ref{fig:example_ttc} and \ref{fig:example_angle} show different calculated variables for the same experiment.

Figs. \ref{fig:example_interacion_exterior} and \ref{fig:example_interaction_invehicle} show the four main states of the pedestrian during an interaction with the AV. First, the pedestrian is standing back to the crosswalk (fig \ref{fig_first_casea}). Then, the pedestrian turns around (fig \ref{fig_first_caseb}) and walks towards the crosswalk (fig \ref{fig_first_casec}) and observes the vehicle approaching (fig \ref{fig_first_cased}). The pedestrian decides if it is safe or not to cross and delays the action if it is not safe enough (figs \ref{fig_first_casee}, and \ref{fig_first_casef}). Finally, the pedestrian feels confident enough to cross through the crosswalk (figs \ref{fig_first_caseg}, and \ref{fig_first_caseg}). 

Fig. \ref{fig:example_interaction_passenger} shows different instances of the passenger interacting with the iHMI while the AV is interacting with the pedestrian at the crosswalk. It can be observed how the iHMI draws the passenger's attention when voice messages are played (figs. \ref{fig_pas_b} and \ref{fig_pas_d}).

Fig. \ref{fig:example_ttc} shows some recorded variables such as the distance to the pedestrian, the speed of the vehicle, and the calculated TTC. On the time axis, there are two time-events marked, one at $t=0$ which corresponds with the labeled crossing event (exemplified in Fig. \ref{fig_first_casef}) and another at $t=-2.3$ approximately, corresponding with the trigger of the braking maneuver.
Fig. \ref{fig:example_angle} shows the recorded variable distance to the pedestrian and the computed solid angle $\Omega(t)$ and its temporal variation $d\Omega(t)/dt$ for the same experiment. Time marks are the same as for Fig. \ref{fig:example_ttc}. By combining the recorded and computed variables with the \textit{crossing event} a set of statistics related to the interaction can be generated. For this example it is known that the pedestrian enters the vehicle lane when the vehicle is at 8 meters distance, driving at 14 kph, representing a 0.5 $sr$ solid angle with a change rate of 0.4 $sr/s$. 

The solid angle represented by the vehicle follows an opposite trend as the distance and the speed. While speed and distance decrease when the vehicle is approaching the pedestrian the solid angle increases. The change rate of the solid angle $d\Omega/dt$ has a different behavior. It is similar to the solid angle $\omega$ at far distances, but as a result of the vehicle deceleration, it starts to decrease while the solid angle continues growing. This inflection point can be observed in Fig. \ref{fig:example_angle}, approximately at $t=1 s$. It can be observed that the \textit{crossing event}, which is a posterior manifestation of the \textit{crossing decision}, is produced 0.8 seconds after the change rate of the solid angle reaches the 0.2 $sr/s$ threshold.

%With this set of variables is easy to understand that the higher the confidence in the car the earlier the \textit{crossing decision} and consequently the \textit{crossing event}, even if the vehicle is driving at higher speeds or closer.


\input{8_table_frequencies}

%% file: 8_table_frequencies.tex
\begin{table*}
\renewcommand{\arraystretch}{1.1}
\setlength{\tabcolsep}{4pt}
\caption{Answers' frequency by test and question}
\begin{center}
\begin{tabular}{cc|cccc|cccc|cccc|cccc|cccc|cccc|cccc}
\multicolumn{1}{l}{}             &   & \multicolumn{4}{c|}{Question 1} & \multicolumn{4}{c|}{Question 2} & \multicolumn{4}{c|}{Question 3} & \multicolumn{4}{c|}{Question 4} & \multicolumn{4}{c|}{Question 5} & \multicolumn{4}{c|}{Question 6} & \multicolumn{4}{c}{Question 7}\\
\multicolumn{2}{l|}{Test n$^\circ$} &1&2&3&4&1&2&3&4&1&2&3&4&1&2&3&4&1&2&3&4&1&2&3&4&1&2&3&4  \\
\hline
\multirow{8}{*}{\rotatebox[origin=c]{90}{Answer code}} 
  & 0 &  0 &  0 &  0 &  0 &  0 &  0 &  0 &  0 & \textbf{32} & \textbf{31} & 2 &  1 &  0 &  0 &  0 &  0 &  0 &  0 &  0 &  0 & \textbf{32} & \textbf{32} &  1 &  1 & \textbf{32} & \textbf{32} &  1 &  1 \\
  & 1 &  0 &  0 &  0 &  0 &  4 &  0 &  3 &  0 &  0 &  0 & 1 &  1 &  0 &  0 &  0 &  0 &  2 &  0 &  0 &  0 &  0 &  0 &  3 &  3 &  0 &  0 &  7 & 10 \\
  & 2 &  1 &  2 &  0 &  1 &  5 &  0 &  9 &  0 &  0 &  0 & 0 &  2 &  0 &  0 &  0 &  0 &  6 &  0 &  3 &  0 &  0 &  0 &  0 &  1 &  0 &  0 &  9 &  9 \\ 
  & 3 &  4 &  4 &  1 &  3 &  9 &  0 &  7 &  3 &  0 &  0 & 1 &  2 &  0 &  2 &  0 &  0 &  4 &  1 &  3 &  0 &  0 &  0 &  1 &  2 &  0 &  0 & \textbf{15} & \textbf{12} \\ 
  & 4 &  5 & \textbf{11} &  1 &  4 & \textbf{14} & \textbf{21} & \textbf{13} & \textbf{18} &  0 &  1 & 2 &  5 &  2 &  9 &  1 &  5 & \textbf{20} &  7 & \textbf{25} & 11 &  0 &  0 &  6 &  7 &  0 &  0 &   &   \\
  & 5 &  7 &  5 &  4 &  7 &  0 & 11 &  0 & 10 &  0 &  0 & 8 &  5 &  9 & \textbf{13} &  6 & \textbf{11} &  0 & \textbf{19} &  1 & \textbf{17} &  0 &  0 & \textbf{10} & \textbf{11} &  0 &  0 &   &   \\
  & 6 &  6 &  5 & 12 & \textbf{11} &  0 &  0 &  0 &  1 &  0 &  0 & \textbf{9} & \textbf{11} & 10 &  3 & 12 & 10 &  0 &  5 &  0 &  3 &  0 &  0 &  8 &  6 &  0 &  0 &   &   \\
  & 7 &  \textbf{9} &  5 & \textbf{14} &  6 &  0 &  0 &  0 &  0 &  0 &  0 & \textbf{9} &  5 & \textbf{11} &  5 & \textbf{13} &  6 &  0 &  0 &  0 &  1 &  0 &  0 &  3 &  1 &  0 &  0 &   &   \\
\end{tabular}
\label{tab:freqs}
\end{center}
\end{table*}

%% file: 4_results.tex
\section{Results}\label{sec:results}
This section presents and analyzes systematically for each type of interaction with the AV the responses to the questionnaires in subsection \ref{subsec:result_questionnaire} and the measured and derived variables in subsection \ref{subsec:result_measures}. Descriptive statistics and the Wilcoxon signed-rank test for paired samples have been used to conduct the analysis of the questionnaires. On the other hand, the Student t-test has been used to extract information from the direct measured and computed variables.

% T-student analysis based on user answers and direct measurements in the experiments.
% Prueba t-student para muestras relacionadas

% Aquí está la trampa con la escala de Likert: muchos investigadores la tratarán como una escala de intervalo. Esto supone que las diferencias entre cada respuesta son iguales en la distancia. La verdad es que la escala de Likert no nos dice eso.

% La prueba de los rangos con signo de Wilcoxon es una prueba no paramétrica para comparar el rango medio de dos muestras relacionadas y determinar si existen diferencias entre ellas. Se utiliza como alternativa a la prueba t de Student cuando no se puede suponer la normalidad de dichas muestras. Wilcoxon, muestras apareadas

\subsection{Questionnaire Results} \label{subsec:result_questionnaire}
This subsection presents the results of the surveys conducted before and during the experiments. These results are summarized in table \ref{tab:freqs} employing the frequency of the response to each of the answers and experiments. Note that for the test-0 the answers have been omitted for representation purposes. For question number 7 the answer code is interpreted as 0 - \textit{not detected}, 1 - \textit{visual}, 2 - \textit{audio}, and 3 - \textit{both}. The mode is represented in Table \ref{tab:freqs} with bold values and the median can be easily computed.

The responses of each participant are now evaluated against tests, by means of the alternative hypothesis matrix. Using the Wilcoxon signed-rank test the answers provided by each participant are evaluated to find differences with statistical significance among the interactions. For the significance level, a parameter $\alpha$=0.05 has been selected. The alternative hypothesis matrix systematically evaluates the null hypothesis of the specific test against others. As the null hypothesis, we propose $H_0:\mu_i\leq\mu_j$ and as the alternative hypothesis, we take $H_1:\mu_i>\mu_j$. A checkmark in a specific cell in the matrix means that $H_0$ is rejected and $H_1$ is accepted when comparing the answers provided in test $i$ (row) with test $j$ (column). In this specific context, the rejection of $H_0$ means that there is a difference with statistical significance between the answers for tests $i$ and $j$, and the answers for test $i$ have a higher score in the Likert scale than for test $j$. 

Table \ref{tab:wilcoxon} shows the alternative hypothesis matrix. Cells with a checkmark represent cases in which the null hypothesis $H_0$ is rejected and the alternative hypothesis $H_1$ is accepted. This table will be used in section \ref{sec:discussion} to interpret the effect of the different test configurations on the participants and their confidence on the AV.

\begin{table}[h]
\renewcommand{\arraystretch}{1.1}
\caption{Wilcoxon Signed Rank test for Questions Q1-Q6}
\begin{center}
\begin{tabular}{ccc|p{1cm}p{1cm}p{1cm}p{1cm}}
%\hline
\multicolumn{3}{c|}{\textbf{$H_1:\mu_i>\mu_j$}}& \multicolumn{4}{c}{\textbf{Test number $j$}} \\
%\cline{3-7} 
 \multicolumn{3}{c|}{} & 1 & 2 & 3 & 4\\
\hline
% & & %\multicolumn{4}{|c|}{\textbf{Question Q1}} \\
%\cline{3-7}
\multirow{24}{*}{\rotatebox[origin=c]{90}{\textbf{Test number $i$}}} & \multirow{4}{*}{\rotatebox[origin=c]{90}{\textbf{Q1}}}     & 1   & --   & \checkmark    &       & \\
                                            &&2   &     & --    &      & \\
                                            &&3   & \checkmark   & \checkmark     & --    & \checkmark\\
                                            &&4   &     & \checkmark     &      & --\\

\cline{3-7}
&\multirow{4}{*}{\rotatebox[origin=c]{90}{\textbf{Q2}}} & 1   & --   &      &      & \\
                                                        &&2   & \checkmark    & --    & \checkmark     & \\
                                                        &&3   &     &      & --    & \\
                                                        &&4   & \checkmark    &      & \checkmark     & --    \\

\cline{3-7}
&\multirow{4}{*}{\rotatebox[origin=c]{90}{\textbf{Q3}}} & 1   & --   &      &      & \\
                                                        &&2   &     & --    &      &      \\
                                                        &&3   & \checkmark    & \checkmark     & --    & \checkmark     \\
                                                        &&4   & \checkmark    & \checkmark     &      & --    \\

\cline{3-7}
&\multirow{4}{*}{\rotatebox[origin=c]{90}{\textbf{Q4}}} & 1   & --   & \checkmark     &      & \checkmark\\
                                                        &&2   &     & --    &      &      \\
                                                        &&3   &     & \checkmark     & --    & \checkmark     \\
                                                        &&4   &     & \checkmark     &      & --    \\

\cline{3-7}
&\multirow{4}{*}{\rotatebox[origin=c]{90}{\textbf{Q5}}} & 1   & --   &      &      & \\
                                                        &&2   & \checkmark    & --    & \checkmark     &      \\
                                                        &&3   & \checkmark    &      & --    &      \\
                                                        &&4   & \checkmark    &      & \checkmark     & --    \\

\cline{3-7}
&\multirow{4}{*}{\rotatebox[origin=c]{90}{\textbf{Q6}}} & 1   & --   &      &      & \\
                                                        &&2   &     & --    &      &      \\
                                                        &&3   & \checkmark    & \checkmark     & --    &      \\
                                                        &&4   & \checkmark    & \checkmark     &      & --    \\
\end{tabular}
\label{tab:wilcoxon}
\end{center}
\end{table}

\begin{table}[h]
\renewcommand{\arraystretch}{1.2}
\caption{Wilcoxon Signed Rank test for Questions begin first Passenger vs Pedestrian}
\begin{center}
\begin{tabular}{cc|p{1cm}p{1cm}p{1cm}p{1cm}}
%\hline
\multicolumn{2}{c|}{\textbf{$H_1:\mu_{pass}>\mu_{ped}$}}& \multicolumn{4}{c}{\textbf{Test number $j$}} \\
%\cline{3-7} 
 \multicolumn{2}{c|}{} & 1 & 2 & 3 & 4\\
\hline
% & & %\multicolumn{4}{|c|}{\textbf{Question Q1}} \\
%\cline{3-7}
\multirow{6}{*}{\rotatebox[origin=c]{90}{\textbf{Question number}}} 
                                            &1   &      & \checkmark    &       &            \\
                                            &2   &      &               &       &            \\
                                            &3   &      &               &       & \checkmark \\
                                            &4   &      &               &       &            \\
                                            &5   &      &               &       &            \\
                                            &6   &      &               &       &            \\
\end{tabular}
\label{tab:wilcoxonpassenger}
\end{center}
\end{table}

Another investigation is whether participating first as a passenger and then as a pedestrian or vice versa has any effect on the interaction experienced with the AV. Table \ref{tab:wilcoxonpassenger} follows the same alternative hypothesis matrix analysis of the responses provided by the subset of participants being first passengers and the subset of participants being first pedestrians. No consistent differences have been found between these two groups. Only two different questions in two different experiments show differences with statistical significance in their responses.

The previous and after-experimentation questionnaire shows the change in general confidence when interacting with the AV as a pedestrian and passenger. Tables \ref{tab:qba3} and \ref{tab:qba4} show the transition matrix for the responses to questions QBA3 and QBA4, respectively. The red area of the table represents transitions in which the answer has a higher value after the experimentation than before. The blue area represents a lower value for the answer after the experimentation and the green one represents no change in the answer. It can be observed that the general confidence as a user of an AV has been increased after the experimentation (Tab. \ref{tab:qba3}) with 18 increases in the confidence versus 1 decrease. The confidence interacting with an AV as a pedestrian has also increased with 24 responses with a higher confidence value after the experimentation versus 1 with a lower confidence (Tab. \ref{tab:qba4}).

\begin{table}[h]
\renewcommand{\arraystretch}{1.2}
\caption{QBA3 - Answer Transition Matrix}
\begin{center}
\begin{tabular}{cc|ccccccc}
%\hline
\multicolumn{2}{l|}{} & \multicolumn{7}{c}{\textbf{Answer Code After}} \\
\multicolumn{2}{l|}{} & 1 & 2 & 3 & 4 & 5 & 6 & 7\\
\hline
\multirow{7}{*}{\rotatebox[origin=c]{90}{\textbf{Answer Code Before}}} & 
     1  & \cellcolor{green!10} & \cellcolor{red!10} & \cellcolor{red!10} & \cellcolor{red!10} & \cellcolor{red!10} & \cellcolor{red!10} &  \cellcolor{red!10}\\
   & 2  & \cellcolor{blue!10} & \cellcolor{green!10} & \cellcolor{red!10} & \cellcolor{red!10} & \cellcolor{red!10} & \cellcolor{red!10} &  \cellcolor{red!10}\\
   & 3  & \cellcolor{blue!10} & \cellcolor{blue!10} & \cellcolor{green!10} & \cellcolor{red!10} \textcolor{red}{1}& \cellcolor{red!10} \textcolor{red}{2}& \cellcolor{red!10} &  \cellcolor{red!10}\\
   & 4  & \cellcolor{blue!10} & \cellcolor{blue!10} & \cellcolor{blue!10} & \cellcolor{green!10} \textcolor{green}{4}& \cellcolor{red!10} \textcolor{red}{8}& \cellcolor{red!10}\textcolor{red}{2} &  \cellcolor{red!10}\\
   & 5  & \cellcolor{blue!10} & \cellcolor{blue!10} & \cellcolor{blue!10} & \cellcolor{blue!10} & \cellcolor{green!10} \textcolor{green}{4}& \cellcolor{red!10} \textcolor{red}{5}&  \cellcolor{red!10}\\
   & 6  & \cellcolor{blue!10} & \cellcolor{blue!10} & \cellcolor{blue!10} & \cellcolor{blue!10} & \cellcolor{blue!10} \textcolor{blue}{1}& \cellcolor{green!10}\textcolor{green}{5} &  \cellcolor{red!10}\\
   & 7  & \cellcolor{blue!10} & \cellcolor{blue!10} & \cellcolor{blue!10} & \cellcolor{blue!10} & \cellcolor{blue!10} & \cellcolor{blue!10} &  \cellcolor{green!10}\\

\end{tabular}
\label{tab:qba3}
\end{center}
\end{table}

\begin{table}[h]
\renewcommand{\arraystretch}{1.2}
\caption{QBA4 - Answer Transition Matrix}
\begin{center}
\begin{tabular}{cc|ccccccc}
%\hline
\multicolumn{2}{l|}{} & \multicolumn{7}{c}{\textbf{Answer Code After}} \\
\multicolumn{2}{l|}{} & 1 & 2 & 3 & 4 & 5 & 6 & 7\\
\hline
\multirow{7}{*}{\rotatebox[origin=c]{90}{\textbf{Answer Code Before}}} & 
     1  & \cellcolor{green!10} & \cellcolor{red!10} & \cellcolor{red!10} & \cellcolor{red!10} \textcolor{red}{1}& \cellcolor{red!10} & \cellcolor{red!10} &  \cellcolor{red!10}\\
   & 2  & \cellcolor{blue!10} & \cellcolor{green!10} & \cellcolor{red!10} & \cellcolor{red!10} & \cellcolor{red!10} & \cellcolor{red!10} &  \cellcolor{red!10}\\
   & 3  & \cellcolor{blue!10} & \cellcolor{blue!10} & \cellcolor{green!10} & \cellcolor{red!10} \textcolor{red}{4}& \cellcolor{red!10} \textcolor{red}{2}& \cellcolor{red!10} \textcolor{red}{1}&  \cellcolor{red!10}\\
   & 4  & \cellcolor{blue!10} & \cellcolor{blue!10} & \cellcolor{blue!10} & \cellcolor{green!10} \textcolor{green}{3}& \cellcolor{red!10} \textcolor{red}{6}& \cellcolor{red!10} \textcolor{red}{4}&  \cellcolor{red!10}\\
   & 5  & \cellcolor{blue!10} & \cellcolor{blue!10} & \cellcolor{blue!10} & \cellcolor{blue!10} & \cellcolor{green!10} \textcolor{green}{1}& \cellcolor{red!10} \textcolor{red}{6}&  \cellcolor{red!10}\\
   & 6  & \cellcolor{blue!10} & \cellcolor{blue!10} & \cellcolor{blue!10} & \cellcolor{blue!10} & \cellcolor{blue!10} \textcolor{blue}{1}& \cellcolor{green!10}\textcolor{green}{3} &  \cellcolor{red!10}\\
   & 7  & \cellcolor{blue!10} & \cellcolor{blue!10} & \cellcolor{blue!10} & \cellcolor{blue!10} & \cellcolor{blue!10} & \cellcolor{blue!10} &  \cellcolor{green!10}\\

\end{tabular}
\label{tab:qba4}
\end{center}
\end{table}

\subsection{Measurements Results} \label{subsec:result_measures}
This subsection analyzes the direct measures recorded during the experimentation and the variables computed from them. The recorded measures are the position and the speed of the AV which endows the calculation of the distance to the pedestrian, the TTC, and the solid angle and its change rate. 

In contrast to the questionnaires, these variables are analyzed using the Student-t test and the alternative hypothesis matrix. The confidence parameter is also set to $\alpha=$0,05. Table \ref{tab:student} shows the systematical analysis of the variables along experiments.

% \begin{equation}
% t = \dfrac{\overline{X}_1 - \overline{X}_2}{\sqrt{\dfrac{S_{c1}^2}{n1} + \dfrac{S_{c2}^2}{n2}}}
% \end{equation}

\begin{table}[htbp]
\renewcommand{\arraystretch}{1.1}
\caption{Student t-test for Distance, Speed, TTC, $\Omega$ and $d\Omega/dt$ at the Crossing Event}
\begin{center}
\begin{tabular}{ccc|p{1cm}p{1cm}p{1cm}p{1cm}}
%\hline
\multicolumn{3}{c|}{\textbf{$H_1:\mu_i>\mu_j$}}& \multicolumn{4}{c}{\textbf{Test number $j$}} \\
%\cline{3-7} 
 \multicolumn{3}{c|}{} & 1 & 2 & 3 & 4\\
\hline
% & & %\multicolumn{4}{|c|}{\textbf{Question Q1}} \\
%\cline{3-7}
\multirow{20}{*}{\rotatebox[origin=c]{90}{\textbf{Test number $i$}}} & \multirow{4}{*}{\rotatebox[origin=c]{90}{\textbf{Distance}}}   & 1   & --   & \checkmark    &       & \checkmark\\
                                                &&2   &     & --    &      & \\
                                                &&3   & \checkmark   & \checkmark     & --    & \checkmark\\
                                                &&4   &     &      &      & --\\
\cline{3-7}
&\multirow{4}{*}{\rotatebox[origin=c]{90}{\textbf{Speed}}}  & 1   & --   & \checkmark     &      & \checkmark\\
                                                            &&2   &     & --    &      &      \\
                                                            &&3   & \checkmark    & \checkmark     & --    & \checkmark     \\
                                                            &&4   &     &      &      & --    \\
\cline{3-7}
&\multirow{4}{*}{\rotatebox[origin=c]{90}{\textbf{TTC}}}    & 1   & --   & \checkmark     &      & \checkmark\\
                                                            &&2   &     & --    &      & \checkmark     \\
                                                            &&3   & \checkmark    & \checkmark     & --    & \checkmark     \\
                                                            &&4   &     &      &      & --    \\
\cline{3-7}
&\multirow{4}{*}{\rotatebox[origin=c]{90}{$\boldsymbol{\Omega}$}}    & 1   & --   &      &  \checkmark    & \\
                                                            &&2   &  \checkmark   & --    &   \checkmark   &      \\
                                                            &&3   &     &      & --    &      \\
                                                            &&4   & \checkmark    &      &  \checkmark    & --    \\
\cline{3-7}
&\multirow{4}{*}{\rotatebox[origin=c]{90}{$\boldsymbol{d\Omega/dt}$}}    & 1   & --   &      & \checkmark     & \\
                                                            &&2   & \checkmark    & --    & \checkmark     &      \\
                                                            &&3   &     &      & --    &      \\
                                                            &&4   & \checkmark    & \checkmark     &  \checkmark    & --    \\
\end{tabular}
\label{tab:student}
\end{center}
\end{table}

There is a special consideration in table \ref{tab:student}. The distance, speed, and TTC are decreasing monotonic variables, at least until the \textit{crossing event} in most of the cases. The TTC starts to grow at a specific point that can take place before or after the \textit{crossing event}. However, the solid angle is an increasing monotonic variable, and its change rate is also an increasing monotonic variable until a point that usually takes place after the \textit{crossing event}. This opposite behavior of the study variables produces that the responses have the opposite difference and the alternative hypothesis matrix shows complementary results for these two variables. It can be observed clearly at the column of test number 3 in table \ref{tab:student}. None of the null hypotheses are rejected in the column of test number 3 according to the distance, speed, or TTC, but it is rejected according to the solid angle and its change rate. 

Following the same structure for the analysis as in \ref{subsec:result_questionnaire}, the alternative hypothesis matrix is also studied for the subset of participants being first pedestrians and passengers. There is no difference in the observed or computed variables between the two groups. For simplicity, this matrix has been deliberately omitted because it is populated only with zeroes.

%% file: 5_discussion.tex
\section{Discussion}\label{sec:discussion}
%INTRO HERE --- EXPAND DISCUSSION BASED ON MORE TABLES
This section analyzes and discusses how the different ways of communication can affect the level of perceived confidence of pedestrians and passengers when interacting with the AV based on the results provided in section \ref{sec:results}.

Based on the results derived from the questionnaire (table \ref{tab:wilcoxon}) we can make the following statements:
\begin{itemize}
    \item \textbf{The optimal braking maneuver increases the pedestrian's confidence in the AV}. This statement is based on the analysis of question 1 comparing the test using the optimal braking maneuver (tests 1 and 3) versus the test using the aggressive maneuver (tests 2 and 4). For the sake of the document, these comparisons are abbreviated using the following notation (Q1: t1 vs t2 and t3 vs t4), where Q1 makes reference to the question that supports the statement and the number of the tests whether the variation is included or not.
    \item \textbf{The eHMI increases the pedestrian's confidence on the vehicle} (Q1: t3 vs t1 and t4 vs t2). %Based on responses given to question 1 comparing the tests using the eHMI (tests 3 and 4) versus the test without the eHMI (tests 1 and 2) %(Q1: t3 vs t1 and t4 vs t2)
    \item \textbf{Pedestrians perceived the aggressive braking maneuvers as “more aggressive” than the gentle breaking maneuvers} (Q2: t2 vs t1 and t4 vs t3). This question is a manipulation check that verifies that the aggressive and optimal braking maneuvers are perceived in such a way.
    \item The optimal braking maneuver increases the passenger's confidence in the AV (Q4: t1 vs t2 and t3 vs t4).
    \item \textbf{The iHMI increases the passenger's confidence in the vehicle when stopping with an aggressive braking maneuver} (Q4: t4 vs t2).
    \item The data does not present differences with statistical significance to support that the iHMI increases the passenger's confidence in the vehicle when stopping in a crosswalk using an optimal braking maneuver (Q4: t3 vs t1). The data suggests that in the case of smooth driving behavior, there is no risk perception and consequently the use of the iHM does not increase the passenger's confidence.
    \item {Passengers perceived the aggressive braking maneuvers as “more aggressive” than the gentle braking maneuvers} (Q5: t2 vs t1 and t4 vs t3). The manipulation check verifies the perception of the gentle and aggressive braking maneuvers.
    \item \textbf{Passengers preferred the combined mode (audio plus video) for the iHMI rather than the audio or video choices for the iHMI} based on the answer's frequency for Q7.
\end{itemize}

Results derived from the direct measures (table \ref{tab:student}) support the following statements:
\begin{itemize}
    \item \textbf{The optimal braking maneuver increases the distance at the crossing event} (dist.: t1 vs t2 and t3 vs t4), equivalently decreasing the solid angle and its change rate. Higher distances at the crossing event imply an earlier crossing decision. This statement derived from direct measures reinforces the statement derived from the questionnaire.   
    \item \textbf{The eHMI in combination with the optimal braking maneuver increases the distance at the crossing event} (dist.: t3 vs t1). In opposition to the results derived from the questionnaire the distance at the crossing event only increases when the eHMI is used in combination with the gentle breaking maneuver but not with the aggressive one (dist.: t4 vs t2). This difference between the responses and the observed behavior suggests that in case of real danger, using the eHMI does not effectively anticipate the crossing decision or the crossing event despite the increase in perceived confidence in the AV.
\end{itemize}

%We cannot state that the eHMI contributes to increasing the distance at the crossing event when the AV brakes with the aggressive braking maneuver (dist.: t4 vs t2). This suggests that subjectively pedestrians feel safer (based on their own responses) but objectively they are not and the perception of risk dominates their actual behavior. 

As an additional finding, there were no statistical differences between the responses and measures obtained from the participants regardless of their initial role played.

Attending to the changes in the answers to question QBA3: 18 (56.25\%) participants expressed a higher confidence regarding the use of an AV after the experiments, 13 (37.5\%) the same confidence, and 1 (3.125\%) a lower confidence.
Attending to the changes in the answers to question QBA4: 24 (75\%) participants expressed a higher confidence interacting with an AV as a pedestrian after the experiments, 7 (21.88\%) the same confidence, and 1 (3.125\%) a lower confidence.

%% file: 6_conclusions.tex
\section{Conclusions and Future Work}
This work explores the capabilities of explicit and implicit ways of communication to affect the confidence of passengers and pedestrians in autonomous vehicles when interacting in a crosswalk through real-world experimentation. Two different braking speed profiles in combination with the use (or not) of internal and external HMIs have been evaluated under these experiments. Questionnaires related to the user's experience during the interaction and direct measures such as the distance at the crossing event have been used to extract conclusions from the experiments.

Questionnaires and direct measurements have proven that the iHMI and eHMI in combination with a gentle braking maneuver help to increase the confidence in the AV when interacting with a pedestrian in a crosswalk for both the pedestrian and the passenger. However, there is a relevant difference between conclusions derived from questionnaires and measured variables. When comparing the experiments using the aggressive braking maneuver, pedestrians express more confidence when using the eHMI than when not using it. However, it does not result in an earlier crossing event and consequently in an earlier crossing decision. This fact suggests that the perception of risk due to the vehicle dynamics has more weight in the decision than the information shared from the eHMI.

As a future work, this study can be extended by including interactions in non-signalized crossing areas where the pedestrian has no priority over the incoming vehicle. The study can also be extended by including age groups not present in the current sample, such as children and teenagers, as part of the collective that will interact with the AVs. Due to the difficulty of performing this study in a controlled real-world environment and the limitation to extending this study to children, a possible solution could be its replication in Virtual Reality (VR) not only to extend the population target of the study but also to measure the reality gap among the users already present in the current study.

%% file: 8_anexo_I.tex
\textbf{- QBA1}: What is your knowledge about autonomous vehicles?

\textit{1) None; 2) Very little; 3) Little; 4) Medium; 5) Quite a lot; 6) A lot; 7) Expert.}
\vspace{2.0mm}

\textbf{- QBA2}: Have you had any experience as a user or pedestrian with an autonomous vehicle?

\textit{1) Yes; 2) No; 3) Do not know / Do not answer.}
\vspace{2.0mm}

\textbf{- QBA3}: What is your confidence regarding the use of an autonomous vehicle?

\textit{1) Not at all; 2) Very little; 3) Little; 4) Medium; 5) Quite a lot; 6) A lot; 7) Total.}
\vspace{2.0mm}

\textbf{- QBA4}: As a pedestrian, what is your confidence regarding interaction with an autonomous vehicle?

\textit{1) Not at all; 2) Very little; 3) Little; 4) Medium; 5) Quite a lot; 6) A lot; 7) Total.}
\vspace{2.0mm}

% Hasta aqui, esta see deja fuera para el proximo artículo
% \item[--] Given the event of an autonomous vehicle driving and a pedestrian crossing in an area without a crosswalk: how much do you agree with the following statements? 
% \begin{itemize}
% \item[--] QBA5: The autonomous vehicle should not stop. If there is a collision it would be the pedestrian's fault.
% \begin{enumerate}
% \item Strongly disagree
% \item Disagree
% \item Partially disagree
% \item Neutral
% \item Partially agree
% \item Agree
% \item Strongly agree
% \end{enumerate}

% \item[--] QBA6: The autonomous vehicle should always stop, regardless of whether the pedestrian crosses improperly.
% \begin{enumerate}
% \item Strongly disagree
% \item Disagree
% \item Partially disagree
% \item Neutral
% \item Partially agree
% \item Agree
% \item Strongly agree
% \end{enumerate}

% \item[--] QBA7: The autonomous vehicle should take actions (e.g., reduce its speed, communicate with some visual or audio interface,  etc.) to dissuade the pedestrian from crossing.
% \begin{enumerate}
% \item Strongly disagree
% \item Disagree
% \item Partially disagree
% \item Neutral
% \item Partially agree
% \item Agree
% \item Strongly agree
% \end{enumerate}
% \end{itemize}

%\end{itemize}

The prefix QBA indicates that the questions were answered both before (QB) and after (QA) conducting the experiments.

%% file: 8_anexo_II.tex
When participating as a pedestrian the questions are:

\textbf{- Q1}: What was your level of confidence that the vehicle would stop and yield to you?

\textit{1) No confidence; 2) Very little confidence; 3) Little confidence; 4) Medium confidence; 5) Quite a lot of confidence; 6) A lot of confidence; 7) Total confidence.}
\vspace{2.0mm}

\textbf{- Q2}: How did you perceive the braking of the vehicle?

\textit{1) Too conservative; 2) Quite conservative; 3) Somewhat conservative; 4) Adequate; 5) Somewhat aggressive; 6) Quite aggressive; 7) Too aggressive.}
\vspace{2.0mm}

\textbf{- Q3}: Has the visual communication interface improved your confidence to cross?

\textit{0) Do not perceive any visual signal; 1) Not at all; 2) Very little; 3) A little; 4) Somewhat; 5) Quite a lot; 6) A lot; 7) Very much.}
\vspace{2.0mm}

When participating as a passenger the questions are:

\textbf{- Q4}: What has been your confidence in the vehicle?

\textit{1) No confidence; 2) Very little confidence; 3) Little confidence; 4) Medium confidence; 5) Quite a lot of confidence; 6) A lot of confidence; 7) Total confidence.}
\vspace{2.0mm}

\textbf{- Q5}: How did you perceive the braking of the vehicle?

\textit{1) Too conservative; 2) Quite conservative; 3) Somewhat conservative; 4) Adequate; 5) Somewhat aggressive; 6) Quite aggressive; 7) Too aggressive.}
\vspace{2.0mm}

\textbf{- Q6}: Has the audiovisual communication interface improved the level of confidence in the vehicle?

\textit{0) Do not perceive any visual signal; 1) Not at all; 2) Very little; 3) A little; 4) Somewhat; 5) Quite a lot; 6) A lot; 7) Very much.}
\vspace{2.0mm}

\textbf{- Q7}: Which signal was most helpful to you?

\textit{0) iHMI not detected; V) Visual; A) Audio; B) Both.}
\vspace{2.0mm}

%% file: 7_bio.tex
\vspace{11pt}

\vspace{11pt}
\vfill

%% file: bare_jrnl_new_sample4.bbl
\begin{thebibliography}{10}
\providecommand{\url}[1]{#1}
\csname url@samestyle\endcsname
\providecommand{\newblock}{\relax}
\providecommand{\bibinfo}[2]{#2}
\providecommand{\BIBentrySTDinterwordspacing}{\spaceskip=0pt\relax}
\providecommand{\BIBentryALTinterwordstretchfactor}{4}
\providecommand{\BIBentryALTinterwordspacing}{\spaceskip=\fontdimen2\font plus
\BIBentryALTinterwordstretchfactor\fontdimen3\font minus
  \fontdimen4\font\relax}
\providecommand{\BIBforeignlanguage}[2]{{%
\expandafter\ifx\csname l@#1\endcsname\relax
\typeout{** WARNING: IEEEtran.bst: No hyphenation pattern has been}%
\typeout{** loaded for the language `#1'. Using the pattern for}%
\typeout{** the default language instead.}%
\else
\language=\csname l@#1\endcsname
\fi
#2}}
\providecommand{\BIBdecl}{\relax}
\BIBdecl

\bibitem{Li2019}
M.~Li, B.~E. Holthausen, R.~E. Stuck, and B.~N. Walker, ``No risk no trust:
  Investigating perceived risk in highly automated driving,'' in
  \emph{Proceedings of the 11th AutomotiveUI Conference}.\hskip 1em plus 0.5em
  minus 0.4em\relax Association for Computing Machinery, 2019, p. 177–185.

\bibitem{Llorca2023}
D.~Fernández-Llorca and E.~Gómez, ``Trustworthy artificial intelligence
  requirements in the autonomous driving domain,'' \emph{Computer}, vol.~56,
  no.~2, pp. 29--39, 2023.

\bibitem{Detjen2021}
H.~Detjen, S.~Faltaous, B.~Pfleging, S.~Geisler, and S.~Schneegass, ``How to
  increase automated vehicles' acceptance through in-vehicle interaction
  design: A review,'' \emph{International Journal of Human-Computer
  Interaction}, vol.~37, no.~4, pp. 308--330, 2021.

\bibitem{Rasouli2020}
A.~{Rasouli} and J.~K. {Tsotsos}, ``Autonomous vehicles that interact with
  pedestrians: A survey of theory and practice,'' \emph{IEEE Transactions on
  Intelligent Transportation Systems}, vol.~21, no.~3, pp. 900--918, 2020.

\bibitem{ignacio_v2v}
I.~Parra, H.~Corrales, N.~Hern{\'a}ndez, S.~Vigre, D.~Llorca, and M.~A. Sotelo,
  ``Performance analysis of vehicle-to-vehicle communications for critical
  tasks in autonomous driving,'' in \emph{2019 IEEE Intelligent Transportation
  Systems Conference (ITSC)}, 2019, pp. 195--200.

\bibitem{dey2021communicating}
D.~Dey, A.~Matviienko, M.~Berger, B.~Pfleging, M.~Martens, and J.~Terken,
  ``Communicating the intention of an automated vehicle to pedestrians: The
  contributions of ehmi and vehicle behavior,'' \emph{it-Information
  Technology}, vol.~63, no.~2, pp. 123--141, 2021.

\bibitem{Martin2023}
S.~Martín, R.~Izquierdo, I.~García~Daza, M.~A. Sotelo, and
  D.~Fernández~Llorca, ``Digital twin in virtual reality for human-vehicle
  interactions in the context of autonomous driving,'' in \emph{IEEE
  Intelligent Transportation Systems Conference}, 2023.

\bibitem{Lagstrom2015}
T.~Lagstrom and V.~M. Lundgren, \emph{AVIP-autonomous vehicles interaction with
  pedestrians}.\hskip 1em plus 0.5em minus 0.4em\relax Gothenborg, Sweden: M.S.
  thesis, Dept. Product-Prod. Develop., 2015.

\bibitem{Izquierdo2023}
R.~Izquierdo, S.~Martín, J.~Alonso, I.~Parra, M.~A. Sotelo, and
  D.~Fernández~Llorca, ``Human-vehicle interaction for autonomous vehicles in
  crosswalk scenarios: Field experiments with pedestrians and passengers,'' in
  \emph{IEEE Intelligent Transportation Systems Conference}, 2023.

\bibitem{drivertive2018}
I.~Parra, R.~Izquierdo, J.~Alonso, A.~García-Morcillo, D.~Fernández-Llorca,
  and M.~A. Sotelo, ``{The Experience of DRIVERTIVE-DRIVERless cooperaTIve
  VEhicle-Team in the 2016 GCDC},'' \emph{IEEE Transactions on Intelligent
  Transportation Systems}, vol.~19, no.~4, pp. 1322--1334, 2017.

\bibitem{berkery2004future}
N.~Berkery and R.~Shoji, ``Future trends in the rear-seat entertainment system
  market,'' SAE Technical Paper, Tech. Rep., 2004.

\bibitem{SAE}
S.~Taxonomy, ``Definitions for terms related to driving automation systems for
  on-road motor vehicles (j3016),'' Technical report, Society for Automotive
  Engineering, Tech. Rep., 2016.

\bibitem{engelbrecht2013fahrkomfort}
A.~Engelbrecht, \emph{Fahrkomfort und Fahrspa? bei Einsatz von
  Fahrerassistenzsystemen}.\hskip 1em plus 0.5em minus 0.4em\relax disserta
  Verlag, 2013.

\bibitem{elbanhawi2015passenger}
M.~Elbanhawi, M.~Simic, and R.~Jazar, ``In the passenger seat: investigating
  ride comfort measures in autonomous cars,'' \emph{IEEE Intelligent
  transportation systems magazine}, vol.~7, no.~3, pp. 4--17, 2015.

\bibitem{baker1992discomfort}
C.~F. Baker, ``Discomfort to environmental noise: Heart rate responses of sicu
  patients,'' \emph{Critical Care Nursing Quarterly}, vol.~15, no.~2, p.~75,
  1992.

\bibitem{allsop2017eye}
J.~Allsop, R.~Gray, H.~H. B{\"u}lthoff, and L.~Chuang, ``Eye movement planning
  on single-sensor-single-indicator displays is vulnerable to user anxiety and
  cognitive load.'' \emph{Journal of Eye Movement Research}, vol.~10, no.~5,
  2017.

\bibitem{hartwich2020passenger}
F.~Hartwich, C.~Schmidt, D.~Gr{\"a}fing, and J.~F. Krems, ``In the passenger
  seat: Differences in the perception of human vs. automated vehicle control
  and resulting hmi demands of users,'' in \emph{HCI in Mobility, Transport,
  and Automotive Systems. Automated Driving and In-Vehicle Experience Design:
  Second International Conference, MobiTAS 2020, Held as Part of the 22nd HCI
  International Conference, HCII 2020, Copenhagen, Denmark, July 19--24, 2020,
  Proceedings, Part I 22}.\hskip 1em plus 0.5em minus 0.4em\relax Springer,
  2020, pp. 31--45.

\bibitem{benderius2017best}
O.~Benderius, C.~Berger, and V.~M. Lundgren, ``The best rated human--machine
  interface design for autonomous vehicles in the 2016 grand cooperative
  driving challenge,'' \emph{IEEE Transactions on intelligent transportation
  systems}, vol.~19, no.~4, pp. 1302--1307, 2017.

\bibitem{AlAdawy2019}
D.~AlAdawy, M.~Glazer, J.~Terwilliger, H.~Schmidt, J.~Domeyer, B.~Mehler,
  B.~Reimer, and L.~Fridman, ``Eye contact between pedestrians and drivers,''
  in \emph{10th International Driving Symposium on Human Factors in Driver
  Assessment, Training and Vehicle Design}, 2019, pp. 301--307.

\bibitem{de2019perceived}
M.~{\'A}. de~Miguel, D.~Fuchshuber, A.~Hussein, and C.~Olaverri-Monreal,
  ``Perceived pedestrian safety: Public interaction with driverless vehicles,''
  in \emph{2019 IEEE intelligent vehicles symposium (IV)}.\hskip 1em plus 0.5em
  minus 0.4em\relax IEEE, 2019, pp. 90--95.

\bibitem{bazilinskyy2019survey}
P.~Bazilinskyy, D.~Dodou, and J.~De~Winter, ``Survey on ehmi concepts: The
  effect of text, color, and perspective,'' \emph{Transportation research part
  F: traffic psychology and behaviour}, vol.~67, pp. 175--194, 2019.

\bibitem{chang2022can}
C.-M. Chang, K.~Toda, X.~Gui, S.~H. Seo, and T.~Igarashi, ``Can eyes on a car
  reduce traffic accidents?'' in \emph{Proceedings of the 14th international
  conference on automotive user interfaces and interactive vehicular
  applications}, 2022, pp. 349--359.

\bibitem{modes5}
D.~Moore, R.~Currano, G.~E. Strack, and D.~Sirkin, ``The case for implicit
  external human-machine interfaces for autonomous vehicles,'' in
  \emph{Proceedings of the 11th international conference on automotive user
  interfaces and interactive vehicular applications}, 2019, pp. 295--307.

\bibitem{mason2022lighting}
B.~Mason, S.~Lakshmanan, P.~McAuslan, M.~Waung, and B.~Jia, ``Lighting a path
  for autonomous vehicle communication: the effect of light projection on the
  detection of reversing vehicles by older adult pedestrians,''
  \emph{International journal of environmental research and public health},
  vol.~19, no.~22, p. 14700, 2022.

\bibitem{lynch2016social}
M.~Lynch, ``Social constructivism in science and technology studies,''
  \emph{Human Studies}, vol.~39, pp. 101--112, 2016.

\bibitem{toyotaeHMI}
Toyota, ``Automated driving white paper,'' Toyota Research Institute, Tech.
  Rep., 2022.

\bibitem{rouchitsas2019external}
A.~Rouchitsas and H.~Alm, ``External human--machine interfaces for autonomous
  vehicle-to-pedestrian communication: A review of empirical work,''
  \emph{Frontiers in psychology}, vol.~10, p. 2757, 2019.

\bibitem{habibovic2019external}
A.~Habibovic, J.~Andersson, V.~Malmsten~Lundgren, M.~Klingeg{\aa}rd,
  C.~Englund, and S.~Larsson, ``External vehicle interfaces for communication
  with other road users?'' in \emph{Road Vehicle Automation 5}.\hskip 1em plus
  0.5em minus 0.4em\relax Springer, 2019, pp. 91--102.

\bibitem{heidi_project}
\BIBentryALTinterwordspacing
{HEIDI}, ``Holistic and adaptive interface design for human-technology
  interactions,'' \emph{EU's Horizon Europe research and innovation programme.
  Grant Agreement No. 101069538}, 2023. [Online]. Available:
  \url{https://heidi-project.eu/}
\BIBentrySTDinterwordspacing

\bibitem{wilde1980immediate}
G.~S. Wilde, ``Immediate and delayed social interaction in road user
  behaviour,'' \emph{Applied Psychology}, vol.~29, no.~4, pp. 439--460, 1980.

\bibitem{price2000relationship}
J.~M. Price and S.~J. Glynn, ``The relationship between crash rates and
  drivers' hazard assessments using the connecticut photolog,'' in
  \emph{Proceedings of the Human Factors and Ergonomics Society Annual
  Meeting}, vol.~44, no.~20.\hskip 1em plus 0.5em minus 0.4em\relax SAGE
  Publications Sage CA: Los Angeles, CA, 2000, pp. 3--263.

\bibitem{crundall1999driving}
D.~Crundall, ``Driving experience and the acquisition of visual information,''
  Ph.D. dissertation, University of Nottingham, 1999.

\bibitem{dijksterhuis2015impact}
C.~Dijksterhuis, B.~Lewis-Evans, B.~Jelijs, D.~de~Waard, K.~Brookhuis, and
  O.~Tucha, ``The impact of immediate or delayed feedback on driving behaviour
  in a simulated pay-as-you-drive system,'' \emph{Accident Analysis \&
  Prevention}, vol.~75, pp. 93--104, 2015.

\bibitem{risser1985behavior}
R.~Risser, ``Behavior in traffic conflict situations,'' \emph{Accident Analysis
  \& Prevention}, vol.~17, no.~2, pp. 179--197, 1985.

\bibitem{tom2011gender}
A.~Tom and M.-A. Grani{\'e}, ``Gender differences in pedestrian rule compliance
  and visual search at signalized and unsignalized crossroads,'' \emph{Accident
  Analysis \& Prevention}, vol.~43, no.~5, pp. 1794--1801, 2011.

\bibitem{kotseruba2016joint}
I.~Kotseruba, A.~Rasouli, and J.~K. Tsotsos, ``{Joint attention in autonomous
  driving (JAAD)},'' in \emph{Proceedings of International Conference on
  Computer Vision (ICCV)}, 2017, pp. 206--2013.

\bibitem{dipietro1970pedestrian}
C.~M. DiPietro and L.~E. King, ``Pedestrian gap-acceptance,'' \emph{Highway
  Research Record}, no. 308, 1970.

\bibitem{neale2005overview}
V.~L. Neale, T.~A. Dingus, S.~G. Klauer, J.~Sudweeks, and M.~Goodman, ``An
  overview of the 100-car naturalistic study and findings,'' \emph{National
  Highway Traffic Safety Administration, Paper}, vol.~5, p. 0400, 2005.

\bibitem{rasouli2017agreeing}
A.~Rasouli, I.~Kotseruba, and J.~K. Tsotsos, ``Agreeing to cross: How drivers
  and pedestrians communicate,'' in \emph{2017 IEEE Intelligent Vehicles
  Symposium (IV)}.\hskip 1em plus 0.5em minus 0.4em\relax IEEE, 2017, pp.
  264--269.

\bibitem{sun2003modeling}
D.~Sun, S.~Ukkusuri, R.~F. Benekohal, and S.~T. Waller, ``Modeling of
  motorist-pedestrian interaction at uncontrolled mid-block crosswalks,'' in
  \emph{Transportation Research Record, TRB Annual Meeting CD-ROM, Washington,
  DC}, 2003.

\bibitem{wang2010study}
T.~Wang, J.~Wu, P.~Zheng, and M.~McDonald, ``Study of pedestrians' gap
  acceptance behavior when they jaywalk outside crossing facilities,'' in
  \emph{13th International IEEE Conference on Intelligent Transportation
  Systems}.\hskip 1em plus 0.5em minus 0.4em\relax IEEE, 2010, pp. 1295--1300.

\bibitem{clay1995driver}
D.~Clay, ``Driver attitude and attribution: implications for accident
  prevention,'' Ph.D. dissertation, Cranfield University, 1995.

\bibitem{lagstrom2016avip}
T.~Lagstr{\"o}m and V.~Malmsten~Lundgren, ``Avip-autonomous vehicles'
  interaction with pedestrians-an investigation of pedestrian-driver
  communication and development of a vehicle external interface,'' Ph.D.
  dissertation, Chalmers University of Technology, 2016.

\bibitem{martin2023digital}
S.~Mart{\'\i}n~Serrano, R.~Izquierdo, I.~Garc{\'\i}a~Daza, M.~{\'A}ngel~Sotelo,
  and D.~Fern{\'a}ndez~Llorca, ``Digital twin in virtual reality for
  human-vehicle interactions in the context of autonomous driving,'' in
  \emph{IEEE International Conference on Intelligent Transportation Systems
  (IEEE ITSC)}, 2023.

\bibitem{zou2023pedestrian}
F.~Zou, J.~Ogle, W.~Jin, P.~Gerard, D.~Petty, and A.~Robb, ``Pedestrian
  behavior interacting with autonomous vehicles during unmarked midblock
  multilane crossings: Role of infrastructure design, av operations and
  signaling,'' \emph{arXiv preprint arXiv:2303.17717}, 2023.

\bibitem{prevention2019}
R.~Izquierdo, A.~Quintanar, I.~Parra, D.~Fernández-Llorca, and M.~A. Sotelo,
  ``{The PREVENTION dataset: a novel benchmark for PREdiction of VEhicles
  iNTentIONs},'' in \emph{2019 IEEE Intelligent Transportation Systems
  Conference (ITSC)}, 2019, pp. 3114--3121.

\end{thebibliography}
